\documentclass[useAMS,usenatbib,usegraphicx,referee]{mn2e}
\usepackage[intlimits]{amsmath} 
\usepackage{amsxtra}     
\usepackage{graphicx}
\usepackage{epstopdf}
\usepackage{epsfig}
\usepackage{times}
\usepackage{subfigure}
\usepackage{lineno}  
\usepackage{amsmath}
\usepackage{amssymb}
\def\lesssim{\mathrel{\rlap{\lower4pt\hbox{\hskip1pt$\sim$}}}<}
\def\gtrsim{\mathrel{\rlap{\lower4pt\hbox{\hskip1pt$\sim$}}}>}

\newcommand{\sv}{\langle\sigma v\rangle}
\newcommand{\lsun}{L_{\odot}}
\newcommand{\msun}{M_{\odot}}
\newcommand{\svcanonical}{3\times 10^{26}{{\rm{cm}}^3/{\rm{s}}}}
\def\be{\begin{equation}}
\def\ee{\end{equation}}
\def\lp{\left(}
\def\rp{\right)}

\begin{document}

\title [Dark Stars and Boosted DM Annihilation Rates]{Dark Stars and Boosted Dark Matter Annihilation Rates }
\author[]{Cosmin Ilie$^1$, Katherine Freese$^1$, and Douglas Spolyar$^2$\\
$^1$Michigan Center for Theoretical Physics, Physics Dept.,
Univ. of Michigan, Ann Arbor, MI 48109\\ 
$^2$Center for Particle Astrophysics, 
Fermi National Accelerator Laboratory, Batavia, IL 60510}
\maketitle
\begin{abstract}
Dark Stars (DS) may constitute the first phase of stellar evolution, powered by dark matter (DM) annihilation.
We will investigate here the properties of DS assuming the DM particle has the required properties to explain the excess positron and electron signals in the cosmic rays detected by the PAMELA and FERMI satellites. Any possible DM interpretation of these signals requires exotic DM candidates, with  annihilation cross sections a few orders of magnitude higher than the canonical value required for correct thermal relic abundance for Weakly Interacting Dark Matter candidates; additionally in most models the annihilation must be preferentially to leptons. Secondly, we study the dependence of DS properties on  the concentration parameter of the initial DM density profile of the halos where the first stars are formed.  We
restrict our study to the DM in the star due to simple (vs. extended) adiabatic contraction and minimal (vs. extended) capture;  this
simple study is sufficient to illustrate dependence on the cross section and concentration parameter.
Our basic results are that the final stellar properties, once the star enters the main sequence,
are always roughly the same, regardless of the value of boosted annihilation or concentration parameter in the range between c=2 and c=5: 
stellar mass $\sim 1000 \msun$, luminosity $\sim 10^7 L_\odot$, lifetime $\sim 10^6$ yrs (for the minimal DM models considered here; additional DM would lead to more massive dark stars).
However,  the lifetime, final mass, and final luminosity  of the DS show some dependence on boost factor and concentration parameter
as discussed in the paper.
\end{abstract}

\begin{keywords}
dark matter -- first stars -- stars: formation -- stars: pre-main-sequence
\end{keywords}

\section{Introduction}

\citet{DS2} first considered the effect of DM particles on the first
stars during their formation. The first stars formed when the universe
was about 200 million years old, at $z=10-50$, in $10^6 M_\odot$
haloes consisting of $85\%$ DM and $15\%$ baryons predominantly in the form of H and
He from big bang nucleosynthesis. The canonical example of particle DM
is Weakly Interacting Massive Particles (WIMPs). In many theories
WIMPs are their own antiparticles and annihilate with themselves
wherever the DM density is high.  The first stars are particularly
good sites for annihilation because they form at high redshifts
(density scales as $(1+z)^3$) and in the high density centers of DM
haloes.  \citet{DS2} found that DM annihilation provides a powerful
heat source in the first stars and suggested that the very first
stellar objects might be {\it Dark Stars} (DS), a new phase of stellar
evolution in which the DM -- while only a negligible fraction of the
star's mass -- provides the key power source for the star through DM
heating. Note that the term 'Dark' refers to the power source, not the
appearance of the star. Dark Stars are stars made primarily of hydrogen
and helium with a smattering of dark matter ($<$1\% of the mass
consists of DM); yet they shine due to DM heating.

In subsequent work, \citet{Freese:2008wh} and ~\citet{DSnl} studied the stellar structure of the dark
stars, and found that these objects grow to be large, puffy ($\sim$ 10 A.U.),
bright ($\sim 10^7 L_\odot$), massive (they grow to at least $\sim 500
M_\odot$), and yet cool ($\sim 6000-10,000$K surface temperature) objects.  They
last as long as they are fed by dark matter.  In another paper, \citet{SMDS} considered the possibility of an extended period
of dark matter heating and the consequent growth to Supermassive Dark Stars $>10^5 \msun$.
By contrast, in the standard
case, when DM heating is not included, Population III stars (the standard fusion powered first stars)
\footnote{ Pop. III stars refer to the first stars to form in the universe and are uncontaminated by previous stellar evolution.  They consist only of the elements produced during Big Bang Nucleosynthesis.}
 form by accretion onto a smaller protostar $\sim 10^{-3} \msun$  \citep{OmuNis98} up to
 $\sim 140\msun$ and surface temperatures exceeding $5\times 10^5$K. This higher surface temperature in the standard picture inhibits the accretion of the gas, as various feedback mechanisms become effective at those high temperatures ~\citep{McKeeTan08}. For reviews of first stars formation in the standard scenario, where DM heating is not included  see e.g. ~\citet{BarLoe01,Yoshidaetal03, BroLar04, RipAbe05, Yoshidaetal06}.

WIMP annihilation produces energy at a rate per unit volume 
\begin{equation}
\hat Q_{DM} = n_\chi^2 \langle \sigma v \rangle m_\chi =
\langle \sigma v \rangle \rho_\chi^2/m_\chi ,
\label{eq:Q}
\end{equation}
where $n_\chi$ is the WIMP number density, m$_\chi$ is the WIMP mass,
and $\rho_\chi$ is the WIMP energy density. The final annihilation products
typically are electrons, photons, and neutrinos. The neutrinos escape
the star, while the other annihilation products are trapped in the
dark star, thermalize with the star, and heat it up.  The luminosity
from the DM heating is
\begin{equation}
\label{DMheating}
L_{DM} \sim f_Q \int \hat Q_{DM} dV 
\end{equation}
where $f_Q$ is the fraction of the annihilation energy deposited in the star
(not lost to neutrinos) and $dV$ is the volume element.

The DM heating rate in Dark Stars scales as the square of the WIMP density times the annihilation cross section, as can be seen from Eq.(\ref{eq:Q}).  The WIMP density inside the star adjusts in response to changes in the star's baryonic mass profile, since the gravitational potential well of the star is determined by the baryons. (The DM profile responds to changes in the gravitational potential due to the conservation of adiabatic invariants.)  

In this paper, we investigate the dependence of Dark Star properties on these two quantities: (i) the annihilation cross section and (ii) the density of the halo within which the star forms, as characterized by the concentration parameter. 

For a short list of papers by various other authors that have continued the work of \citet{DS2} and explored the repercussions of DM heating in the first stars see \citet{Iocco2008,DMfs3, DMfs1, DMfs2, DMfs4, Gondolo2010, Ripamonti:2010ab, Sivertsson2010}. Their potential observability has been discussed in \citet{SMDS, Zackrisson2010, Zackrisson2010b}. The possibility that DM annihilation might have effects on {\it{ today's}} stars  was actually considered in the $`80s$ by various authors such as ~\citet{krauss, bouquet, salatisi}. More recently the effect on today's stars has been re-examined under the assumption that DM is made of  WIMPs \citep{moskalenko, scott1, bertone, scott2} or within the hypothesis of inelastic dark matter ~\citep{Hooper2010}.

\subsection{Boosted Leptophilic Annihilation Motivated by PAMELA data}

Recent measurements of cosmic ray positrons and electrons in the GeV-TeV range could have significant implications on our understanding of dark matter (DM). The PAMELA collaboration ~\citep{PAMELA, Adriani2010} reported a $e^{+}$ flux excess in the cosmic energy spectrum from $10$ to $100$ GeV, reinforcing what was previously observed  up to an energy of $50$ GeV by the HEAT experiment ~\citep{HEAT}.  The FERMI-LAT collaboration ~\citep{ Abdo2009, FERMI} has found an excess in the $e^++e^-$ flux in the $100-1000$ GeV energy range above the background given by a conventional diffusive model, albeit in conflict with a much larger excesses in flux in the $500$ GeV range previously reported by ATIC collaboration.\footnote{The ATIC balloon experiment ~\citep{ATIC}  found an increase in the $e^{+} + e^-$ spectrum between $300$ and $800$ GeV above  background which has apparently been ruled out by FERMI.} It is worth mentioning that a simple power-law fit of the FERMI-LAT electron energy spectrum is possible, being consistent with both astrophysical sources, or DM annihilation ~\citep[e.g][]{DiBernardo2009, Grasso2009}. 

If confirmed, there are several possible explanations for the positron excess. The signals could be generated by astrophysical sources, such as pulsars or supernova shocks ~\citep[e.g.][]{boulares, Profumo08, Hooperetal08, Ahlers2009, Blasi09, Fujita2009, Malyshev2009, Mertsch2009, Shaviv2009}. Uncertainties in cosmic ray propagation in the galaxy leave open the possibility that standard cosmic ray physics could explain the signal \citep{delahayesalati}.
Another possible interpretation of the data could be in terms of a signal from DM annihilation ~\citep{Baltz2001, Kane2001, deBoer2002, Hooper2003, Hooper2004, Cirelli2008, Ref:annih3, Grajek2008, B1, nomura, AHDM, Ref:annih1, Barger2009, DMinterpdata, Bai2009, Bi2009, Ref:annih2, Ref:annih5, Ref:annih8, Ref:annih4, Ref:annih7, Ref:annih9, Meade2009, Ref:annih6, Meade2010} or decay ~\citep{decay3, decay1, decay2, Ibarra2008, decay6, decay7, decay4, decay5, B2, Bajc2010, Ref:ChenJCAP10}.  Others point out constraints against such an interpretation \citep{Abdo2010,Abazajian2010}.

The prospect that DM has been detected, although indirectly, has stirred a lot of excitement and a flurry of interest. There are several model-building challenges, however. If dark matter is to explain the positron flux excess in the cosmic-ray spectrum, in most models the products of decay or annihilation must be primarily leptonic since an excess in the anti-proton fluxes is not found ~\citep{PAMELApap} . 

Moreover, in order to explain the observed signals the annihilation rate needs to be on the order of $ 10-10^3$ larger than for thermal relic. This could be explained by non-relativistic enhancements to the cross-section, such as the Sommerfeld enhancement ~\citep{Ref:Somm31, Som3, Som1, Ref:MarchRusselletal08, Ref:annih4, Som2} or a Breit-Wigner enhancement ~\citep{BWreso3, BWreso2, BWreso1, Kadota2010}. Scenarios involving non-standard cosmologies have also been considered as a possible solution ~\citep[e.g][]{Gelmini2008, Catena2009, Pallis2009}. Alternatively, the annihilation rate could be greater than the standard prediction if the dark matter is not smoothly distributed in the local halo of the Milky Way ~\citep{Diemand2008, Hooper2008, Kamionkowski2010}. For a review of the DM explanation for the the cosmic ray $e^{\pm}$ excess see ~\citet{DMreview}.

In light of these new dark matter models, this paper will  study the possible modifications to the evolution  of a Dark Star (DS) for the case of leptophilic boosted dark matter. In \citet{DSnl} a comprehensive study of DS for various WIMP masses was done, but the annihilation cross section was kept to its fiducial unboosted value. We return here to this problem, including the possibility of a boost factor for the cross section. Our starting point will be ~\citet{DMinterpdata}, where the authors find regions in the ($m_{\chi},B$) parameter space  (mass of WIMP vs. boost factor)  that fits  PAMELA/FERMI data based on three different classes of DM models. For simplicity we will consider only two of those models in our study: from the leptophilic class, the $\mu$ channel case, where $100\%$  direct annihilation to $\mu^+\mu^-$ is assumed and the $AH4$ model, a subclass of the Arkani-Hamed model \citep{AHDM} which postulates that the Somerfeld enhancement is due to the exchange of a new type of light (sub- GeV) particle.  In the $AH4$ case the new force carrier is a scalar, $\phi$, which subsequently decays $100\%$ to $\mu^+\mu^-$. Somerfeld enhancement is now generated naturally via ladder diagrams for the $\chi\chi\to \phi\phi$ process, producing 4 muons per annihilation versus 2 muons in the direct annihilation case.

\subsection{Effect of Concentration Parameter}

 A second focus of this paper will be to study the dependence of DS properties on the concentration parameter of the initial density profile of the halo within which the first stars form. To remind the reader, the concentration parameter ($c$) characterizes how centrally condensed the initial density profile is: a larger concentration parameter means that  more of the mass is concentrated at the center rather than at the outside of the dark matter halo (the precise definition is given below in Sec.\ref{secAC}).  Previous studies of DS considered $c=10$  and $c=2$.
 Here we systematically examine a sensible  range  of concentration parameters. The most recent results of numerical  simulations of structure formation seem to favor a "floor" of $c=3.5$ on the concentration parameter of early structures ~\citep{Zhao2003, Zhao2008, Tinker2010} (Some halos can have a smaller concentration parameter if star formation begins before the halo is fully formed).
   At any rate the value will differ from halo to halo, even  at the same redshift.  Thus to  explore how the properties of the DS are affected by a change in the concentration parameter, we will run the same simulation with three different values for it: $c=(2, 3.5, 5)$, which characterizes  the overall range for the  concentration parameter DM halos hosting the first stars.

\subsection{Canonical Values}

We will assume a redshift of formation $z=20$ for the first stars, a dark matter halo of $10^6 \msun$, and concentration parameters 
$c=(2,3.5,5)$. 

 We will take the
annihilation cross section to be 
\begin{equation}
\label{eq:sigmav}
\langle \sigma v \rangle = B \times
3 \times 10^{-26} {\rm cm}^3/{\rm s}
\end{equation}
\noindent  where the boost factor $B$ varies between $100$ and $5000$ (depending on the particle physics model).  The corresponding WIMP mass is taken from the models in \citet{DMinterpdata} described above and ranges from $100$ GeV-$4$ TeV.

A key question for DS is the final mass, as the DS accretes more and more material.  As long as there is a reservoir of DM to heat the DS, the star continues to grow. In the original work of ~\citet{DSnl}, the assumption was made that the initial DM inside the DS annihilates away in $\sim 500,000$ years for a spherical DM halo; here the DS grow to $\sim 1000 \msun$. In later work of \citet{SMDS}, this assumption was questioned due to the fact that DM haloes are instead triaxial, so that a variety of DM orbits can keep the central DM density higher for longer periods of time and the DS can grow supermassive $>10^5 \msun$.   In reality dark stars will form in a variety of dark matter environments and will grow to a variety of masses.  For the purpose of illustrating how DS vary due to differences in the halo concentration parameter and also due to enhanced annihilation rates, we will restrict ourselves to the first option for adiabatic contraction, in which the DM originally in the star (due to adiabatic contraction) is the only DM available to the DS; i.e. the DS can grow to $\sim 1000 \msun$.

 In addition to this simple adiabatic contraction, we will also consider  the effect of captured DM on the first stars (Iocco 2008; Freese, Spolyar,
 and Aguirre 2008). 

In this case, DM passing through the first stars can scatter off the baryons multiple times, lose energy, and become bound to the star (direct detection experiments are based on the same physics: scattering of DM particles off of nuclei).  
 Subsequently, the star builds up a reservoir made up of captured DM, which can power stars.  In the 'minimal capture case,' DM heating from captured DM and fusion powers the star in equal measure once it reaches the main sequence.

In section \ref{sec:setup} we describe the elements necessary to study the stellar structure of the DS.
In Section \ref{sec:results} we present results for
 the influence of varying the concentration parameter and the boost factor on the formation and evolution of Dark Stars and their properties.
 We summarize in Section \ref{sec:summary}.

\section{Equilibrium Structure of the Dark Star}
\label{sec:setup}

\subsection{Initial Conditions and Accretion Rates}\label{inicond}
 The standard picture of Pop. III star formation starts with a protostellar gas cloud that is collapsing and cooling via hydrogen cooling into a protostar \citep{OmuNis98} at the centre of the halo. However, as was found in \citet{DS2},  DM heating could alter the evolution of the first stars significantly. As soon as the protostellar gas reaches a critical core density, DM heating dominates over all possible cooling mechanisms.  The cloud condenses a bit more but then stop collapsing, and becomes a dark star in equilibrium.
  At this point dark matter annihilation can power the dark star.
 
 As the initial conditions for our simulations we take a DS in which the baryons are  fully ionized. This luminous object powered by DM 
 annihilations  has a mass of $3 \msun$, a radius $1-10$ AU, and a central baryon number density $\sim 10^{17}$cm$^{-3}$.  We look for equilibrium solutions as 
 described below. From this starting point we follow the evolution of the DS as it accretes baryonic mass from its surroundings.  We use the 
 accretion rate of ~\citet{Tan2003}, which  decreases from $1.5 \times 10^{-2} \msun/$yr at $3 \msun$ to $1.5\times 10^{-3}\msun/$yr at $1000 
 \msun$. At each stage in the accretion process we again find solutions in hydrostatic and thermal equilibrium.  Eventually the accretion is cutoff 
 by feedback effects; as the dark matter runs out due to annihilation, the DS heats up to the point where it emits ionizing photons that shut down further 
 accretion.  Feedback turns on once the surface temperature $T_{eff}$ reaches 50,000K and is accounted for by introducing a linear reduction factor which shuts off accretion completely once the stars surface temperature reaches $10^5$K.

\subsection{Basic Equations}\label{basiceq}

We use the numerical code previously developed in a paper involving two of the current authors ~\citep{DSnl}. Here we review the ingredients in the code before demonstrating the modifications relevant to this paper.

One key requirement is the hydrostatic equilibrium of the star. This is imposed at each time-step during the accretion process:

\be\label{hydro}
\frac{dP}{dr}=-\rho(r)\frac{GM_r}{r^2}.
\ee
Above $P$ denotes the pressure, $\rho(r)$ is the total density and $M_r$ is the mass enclosed in a spherical shell of radius $r$. We assume that the star can be described as a polytrope with
\be\label{poly}
P=K\rho^{1+1/n},
\ee
where the ``constant'' K is determined once we know the total mass and radius ~\citep{chandra}. 

The  energy transport is initially convective with polytropic index $n=3/2$ but as the star approaches the Zero Age Main Sequence (ZAMS) it becomes radiative with $n=3$. The code interpolates between $n=3/2$ and $n=3$ to account for the shift in energy transport as the star grows in mass. We can determine the temperature at each point in the radial grid via the equation of state of a gas-radiation mixture,
\be\label{eqstate}
P(r)=\frac{k_B\rho(r)T(r)}{m_u\mu}+\frac{1}{3}aT(r)^4\equiv P_g+P_{rad}.
\ee
Here $k_B$ is the Boltzmann constant, $m_u$ is the atomic mass unit and $\mu=(2X+3/4Y)^{-1}$ is the mean atomic weight. In all resulting models $T\gg 10^4$K except near the  surface, so we use the mean atomic weight for fully ionized H and He.  We will assume a H mass fraction of $X=0.76$ and a He mass fraction $Y=0.24$, and that they will remain constant throughout the simulation.

At each point in the radial grid, $T(r)$ and $\rho(r)$ are used to determine the Rosseland mean opacity $\kappa$ from a zero metallicity table from OPAL ~\citep{OPAL} supplemented at low temperatures by opacities from ~\citet{lenzuni} for $T<6000$K.  The location of the photosphere is determined by the hydrostatic condition:
\be\label{photo}
\kappa P=\frac{2}{3}g
\ee
where $g$ is the acceleration due to gravity at that particular location. This corresponds to a point with inward integrated optical depth $\tau\sim2/3$; here the local temperature is set to $T_{eff}$ and the stellar radiated luminosity is therefore:
\be\label{thermal}
L_*=4\pi R_*^2\sigma_BT_{eff}^4,
\ee
with $R_*$ being the photospheric radius.  The thermal equilibrium condition is 
\begin{equation}
L_* = L_{\rm tot}
\end{equation}
where $L_{tot}$ is the total luminosity output from all energy sources as described below in Sec.\ref{ensources}. 

Starting with a mass $M$ and an estimate for the outer radius $R_*$, the code integrates Eqns. (\ref{hydro}) and (\ref{eqstate}) outward from the center.

 The total luminosity output $L_{tot}$ is compared to the stellar radiated luminosity, as in Eq.(\ref{thermal}) and the radius is adjusted until the condition of thermal equilibrium is met (a convergence of $1$ in $10^4$ is reached).  

\subsection{Dark Matter Densities}\label{DMdens}

\subsubsection{Initial Profile and Concentration Parameter}
 The first stars form inside $\sim 10^6 \msun$ haloes.  Simulations imply that DM halos have a naturally cuspy profile, but there is still some uncertainty about the exact inner slope  of a DM 
 halo: ~\citet{2007ApJ...667..859D,2008MNRAS.391.1685S,2010arXiv1002.3660K}.  Luckily, a previous paper ~\citep{DS3}
  showed that a dark star results regardless of the details of the initial density profile, even for the extreme case of a cored Burkert profile (such a Burkert profile is completely unrealistic).  In this paper, we use a Navarro, Frenk, \& White (NFW) profile ~\citep{NFW}  for concreteness. We assume that initially both the baryons (15\% of the mass) and the DM (85\% of the mass) can be described with the same NFW profile

\be \label{iniprofile}
\rho(r)=\frac{\rho_0}{r/r_s(1+r/r_s)^2},
\ee
where $\rho_0$ is the "central" density and $r_s$ is the scale radius.  Clearly at any point of the profile, baryons will only make up 15\% of the mass.
The density scale, $\rho_0$ can be re-expressed in terms of the critical density of the universe at a given redshift, $\rho_c(z)$ via
\be
\rho_0=\rho_c(z)\frac{200}{3}\frac{c^3}{ln(1+c)-c/(c+1)},
\ee 
where $c\equiv r_{vir}/r_s$ is the concentration parameter and $r_{vir}$ is the virial radius of the halo. 
We will assume a flat $\Lambda$CDM universe with current matter density 
$\Omega_m=0.24$ and dark energy density $\Omega_\Lambda=0.76$.

One of the main points of this paper is to study the dependence of DS properties on the value of the
concentration parameter $c$. Hence we will consider a variety of values, ranging from c=2 to c=5.

\subsubsection{Adiabatic Contraction}\label{secAC}
As the baryons start to collapse into a protostellar cloud at the center of the DM halo,
the DM responds to the changing gravitational potential well and becomes compressed.
As described in our previous work, we will use adiabatic contraction (AC) to calculate the effect of baryons on the DM profile. 
 With  an initial DM and gas profile and a final gas profile,  we can use the adiabatic
invariants to solve for the final DM profile.  We use the Blumenthal method~\citep{blumenthal2, blumenthal1, blumenthal3} to calculate the adiabatic compression of the halo.  In this case the simplifying assumption of circular orbits is made. Angular momentum is the only non-zero invariant. Its conservation implies that $M_f(r_f)r_f=M_i(r_i)r_i$.  In the case of 
circular orbits $M$ is the mass interior to the radius $r$ of an orbit and the indices $f$ and $i$ refer to final and initial orbits respectively. As mass grows inside of the orbit, its radius must shrink and the DM profile steepens.

 The validity of this method in this context has been checked in ~\citet{DS3}, where a more precise algorithm, developed by Young ~\citep{young}, has been used and a difference within at most a factor of $2$ has been found. Whereas the Blumenthal method assumes circular orbits, Young's method only assumes  spherical symmetry of the system. Therefore only one of the three conserved actions is identically zero in this case. Namely the plane of each orbit does not change. The other two actions, angular momentum and the radial action respectively are non-zero and conserved. In view of ~\citet{DS3} we are confident that the simple Blumenthal method is sufficiently accurate for our purpose. 
 
 In this paper, we  assume that all of the DM moves on circular orbits.  The DM  will become exhausted  once all of the DM on orbits interior to the DS have been depleted. The timescale for this to happen
 is on the order a million years. This is probably an unduly cautious assumption. In a recent paper ~\citep{SMDS} we studied the case of triaxial haloes with large numbers of centrophilic orbits, which provided a much larger DM reservoir than we are considering, leading to supermassive DS. In reality dark stars will form in a variety of dark matter environments and will grow to a variety of masses. In this paper we could have grown the stars to supermassive sizes, but do not consider it necessary for showing the dependence of DS properties on the two effects we are studying.

\subsubsection{Dark Matter Capture}\label{dmcapture}

Until now we have only discussed the DM inside the DS due
to adiabatic contraction.  However, dark star's DM reservoir can be refueled by DM capture. This refueling requires
an additional piece of particle physics: scattering of DM off the atomic nuclei inside the star.
Some of the  WIMPs from far out in the halo have orbits passing  through the star.  This DM can be captured and sink to the stars center, where they can annihilate efficiently. 
The capture process is irrelevant during the early evolutionary stage of the DS, since the baryon density is not high enough to efficiently capture DM. 
However, at the later stages as the DS contracts towards the ZAMS, the baryon densities
become high enough for substantial capture to be possible.  This mechanism was first
noticed simultaneously by ~\citep{Iocco2008} and ~\citep{DMcap}.

The capture rate is sensitive to two uncertain quantities: the scattering cross section of WIMP interactions with the nuclei and  the background DM density.  
In terms of the relevant particle physics, we will consider only spin-dependent (SD) scattering cross sections with
\be
\sigma_c=10^{-39}{\rm{cm}^2}.
\ee

The details of our capture study have previously been presented in ~\citet{DMcap} and will not be repeated here.  We do wish to emphasize that we will assume the same case of minimal capture that we  previously studied in ~\citet{DSnl}, in which case the heating from fusion and from DM heating are taken to be comparable.  The more extreme and interesting case of dominant capture, which could last as long as the DS continues to exist in a high DM density environment, was studied elsewhere ~\citep{SMDS} and can lead to supermassive $> 10^5 \msun$ DS.

\subsection{Energy Sources}\label{ensources}
There are four possible contributions to the DS luminosity:
\be
L_{tot}=L_{DM}+L_{grav}+L_{nuc}+L_{cap}
\ee
from DM annihilation, gravitational contraction, nuclear fusion, and captured DM respectively.

\subsubsection{DM Annihilation}\label{secdman}
The heating due to DM annihilation is given in Eqns. (\ref{eq:Q}) and (\ref{DMheating}) and dominates
from the time of DS formation until the adiabatically contracted DM runs out.
In order to compute the luminosity generated by DM heating, one needs to know what fraction of the total energy generated by WIMP annihilations is deposited in the star. This quantity, which we name $f_Q$, will be different for various models for DM. In previous work (e.g. in ~\citet{DSnl})   a fiducial value for $f_Q$ of $2/3$ was used as appropriate for many typical WIMPS, under the following assumptions. First,
the final products of DM annihilation, after all unstable particles have decayed to the lightest states, are taken to be three types:
 i) stable charged particles (i.e. $e^\pm$), ii) photons and iii) neutrinos. Second, the energy distribution was assumed to be roughly equal 
 among the three final products listed above.  The electrons and photons were taken to thermalize with the star and deposit their energy, whereas the neutrinos escape; hence $f_Q\simeq 2/3$. 

On the other hand, given a specific model for DM, one could  compute the precise value for $f_{Q}$ using the energy distribution of all final annihilation products for the model under consideration. While this procedure could be done 
for the specific models in this paper, still we will use the standard $f_Q = 2/3$ in order to make simple comparisons with our previous work
\footnote{For the leptophilic models we are considering in this paper, DM annihilates either directly  or via exchange of a light scalar to $100\% \mu^{\pm}$ so that one might more reasonably expect smaller values since the muons decay to electrons and two neutrinos. The actual value will be closer to $\sim 1/3$ but will depend on the energy distribution  among the various final products; 
still it will be a number of $\mathcal{O}(1)$. }.
Since the aim of this paper is to understand the effect of the boost factor on the DS properties, we will fix the value of $f_Q$ to $2/3$, 
the same as in \citet{DSnl}. The differences between the boosted and unboosted cases will now be due only to the boost factor itself and the 
different masses of the WIMP in the two cases (rather than to the detailed values of $f_Q$). Since $f_Q$ appears always multiplied by $\sv$, 
any ambiguity we have introduced by fixing the energy deposition factor to $2/3$ can be traded for an ambiguity in the annihilation cross 
section. 

\subsubsection{Gravitational energy}\label{secgrav}
 Once the DS runs out of DM, it begins to contract; gravity turns on and powers the star. Although relatively short, this Kelvin-Helmholtz (KH) contraction phase has important 
 consequences: it drives up the baryon density and increases the temperature, therefore leading to an environment where nuclear fusion can take place. For a polytrope of index $n=3$ the gravitational contribution to the luminosity was found in ~\citet{DS3} using the virial theorem,
\be\label{lgrav}
L_{grav}=\frac{3}{4}\frac{d}{dt}\lp\frac{GM^2}{R}\rp-\frac{1}{2}\frac{d}{dt}E_{rad},
\ee
where $E_{rad}=\int dVaT^4$ is the radiation energy.

\subsubsection{Nuclear Reactions}
Subsequently nuclear reactions become important.
We include the following nuclear reactions, which are typical for a zero metallicity star during  the pre main sequence evolution: {\it{i}}- burning of primordial deuterium (assumed to have a mass fraction of $4\times 10^{-5}$) which turns on rapidly once the stars central temperature reaches $\sim 10^6$K, {\it{ii}}- the equilibrium proton-proton cycle for hydrogen burning, and {\it{iii}}- the alpha-alpha reaction for helium burning. Since we track the evolution of the DS only until it reaches ZAMS we do not need to consider the changes in stars atomic abundances. In order to calculate the nuclear luminosity, defined as
\be
L_{nuc}=\int dM \epsilon_{nuc}
\ee
we use the methods described in ~\citet{clayton} to obtain the energy generation rate, $\epsilon_{nuc}$. For the proton-proton reaction we use the astrophysical cross section factors from ~\citet{bahcall} and the He burning parameters  from ~\citet{kipp}.

\subsubsection{Luminosity due to captured DM}  
As we have seen in Sec.\ref{dmcapture}, during the later stages of the pre main sequence evolution captured DM can become an important energy source. The luminosity due to DM capture is
\be\label{caplum}
L_{cap}=2m_{\chi}\Gamma_{cap},
\ee
where 
\be\label{gammacap}
\Gamma_{cap}=f_Q\int dV \rho_{cap}^2\sv/m_{\chi}
\ee
and the factor of $2$ in Eq.(\ref{caplum}) reflects the fact that the energy per annihilation is twice the WIMP mass. In all simulations we will consider the case of ``minimal capture'', which corresponds to equal contribution to the luminosity from capture and nuclear fusion when the star reaches the main sequence.   

\section{Results}\label{sec:results}

  On the whole, for all the values of boost factor and concentration parameter we have considered
in this paper, the results are roughly the same: the final DS is roughly $\sim 1000 \msun$, $\sim 10^7 L_\odot$, and lives $\sim 10^6$yrs.  
Thus if the $e^+$ excess seen in PAMELA is due to WIMP annihilation, the required leptophilic boosted cross section is certainly 
compatible with the DS picture.  However
there are interesting differences between models which we will discuss.  

Other than in the subsection immediately following this one, we will consider four WIMP models.  As motivated below, we will focus on
 one boosted model denoted by AH4 with the following set of parameters: $B=1500$, $m_{\chi}=2.35$ TeV and $c=3.5$.  
 As our unboosted
 models, we will take 100 GeV WIMPs with the canonical cross section  of $3 \times 10^{-26}$cm$^3$/s , and we will consider three values of the
 concentration parameter, c = (2, 3.5, 5).  For comparison, the "canonical case" considered in \citet{DSnl} was the unboosted 100 GeV case
 with c=2.  The relative boost factor between the AH4 model and the unboosted models is best described as follows: since Eq.(1) tells us
 that DM heating scales as $\sv/M_\chi$, one can see that the AH4 is exactly equivalent to a 100 GeV WIMP mass with Boost Factor
 $150/2.35\simeq 64$.  In other words the relative boost factor between the AH4 model and the unboosted cases is actually $64$.

 Also, we assume throughout that the boost factor will have  the same  value for $\sv$ that is required to explain the PAMELA/ Fermi data. Particle physics models written down can certainly allow for larger boost factors in the halos in which the first stars form (the converse 
 does not hold;  DM has a lower velocity in a DS halo versus the Milky Way halo, which would act to increase the Sommerfeld enhancement).
 For clarity, we will fix the boost factor to the values given by \citet{DMinterpdata}.

\subsection{`Final' Luminosity for Dark Stars with boosted DM}

We will investigate the `final'  luminosities and masses of  Dark Stars as they enter the main sequence.  We study a variety of WIMP parameter ranges capable of explaining the PAMELA data; in particular we follow the work of ~\citet{DMinterpdata} and take 
 various boosted DM annihilation rates and WIMP masses from fig. 1 or their paper, which   gives $2 \sigma$ contours in the enhancement factor - mass plane needed to fit PAMELA and Fermi data. In this section we consider two of the  particle physics models they
 consider: the $\mu$ channel and AH4 type of models. 

\begin{figure}
\includegraphics[width=0.98\textwidth]{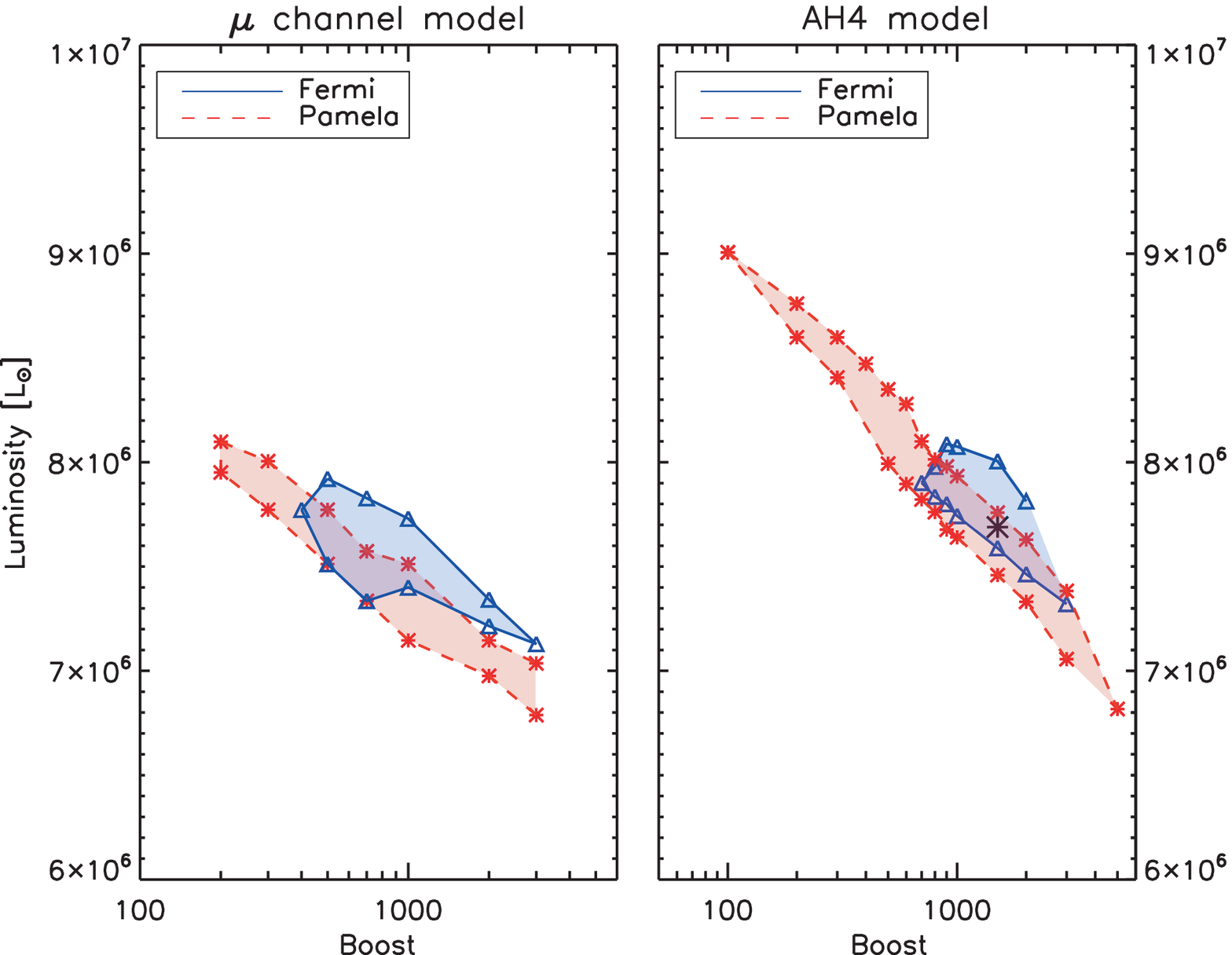}
\caption{``Final'' luminosities (as the Dark Star enters the ZAMS) as a function of  Boost factor for the $\mu$-channel and AH4 models 
examined  in ~\citet{DMinterpdata}.  The contour regions are generated using the range of WIMP masses for a fixed boost factor taken from the corresponding $2 \sigma$ contours in fig. 1 in ~\citet{DMinterpdata}.  The central 
star-shaped point in the right panel will be taken to be our designated boosted AH4 model  and is consistent with both Fermi and PAMELA. }
\label{lvsb}
\end{figure}

In Fig. \ref{lvsb} we plot the `final' luminosity at the end of the DS lifetime, when the DS has accreted to its maximum mass:
the dark matter has run out and the star is about to enter the zero age main sequence where it will be powered by fusion.  We plot the
luminosity as a function of the boost factor. The points represent individual 
simulation outputs. The solid (dashed) lines connect simulation outputs generated with the DM mass and Boost factors from the $2 \sigma$ contours in fig.1 from ~\citet{DMinterpdata} corresponding to the  Fermi (PAMELA) confidence regions in \citet{DMinterpdata}.  We note that even if the boost factors range over more than one order of magnitude, from $100$ to $5000$  ( the corresponding WIMP masses take values in the 1-4 TeV range) the final luminosities in all cases are relatively similar, ranging from $\sim 7\times10^6 L_{\odot}$ to $\sim 9\times10^6 \lsun$  (The variation in luminosity for a fixed boost factor corresponds to the range of allowed $m_\chi$ in~\citet{DMinterpdata}.  For instance for a boost factor of 1000, the upper bound of  the Fermi contour in left panel of Fig. \ref{lvsb} corresponds to a $m_\chi=1.4$~TeV; the lower bound to $m_\chi=1.9$~TeV ). As the boost factor increases, we note that the final luminosity gets slightly smaller. This is due to the fact that the ratio  $\sv/M_{DM}$ typically also increases in most cases. This leads to a faster depletion of the adiabatically contracted DM from the star, shortening the time until it reaches the main sequence, therefore reducing the amount of mass accreted and consequently the luminosity at that point.

\subsection{Structure and Evolution of DS}  
In this section, we will analyze  the two effects of i)  boosting the annihilation cross section and ii) a variety of concentration parameters
on the structure of a Dark Star. As mentioned above, for the ``boosted'' case we choose one sample point from the AH4 type models, 
which corresponds to the large star shaped point at the center of the right panel of Fig. \ref{lvsb}.  Henceforth  we will denote by AH4 this 
point which has $B=1500$, $m_{\chi}=2.35$ TeV and $c=3.5$.

\subsubsection{Luminosity Evolution}

\begin{figure}
\includegraphics[width=0.98\textwidth]{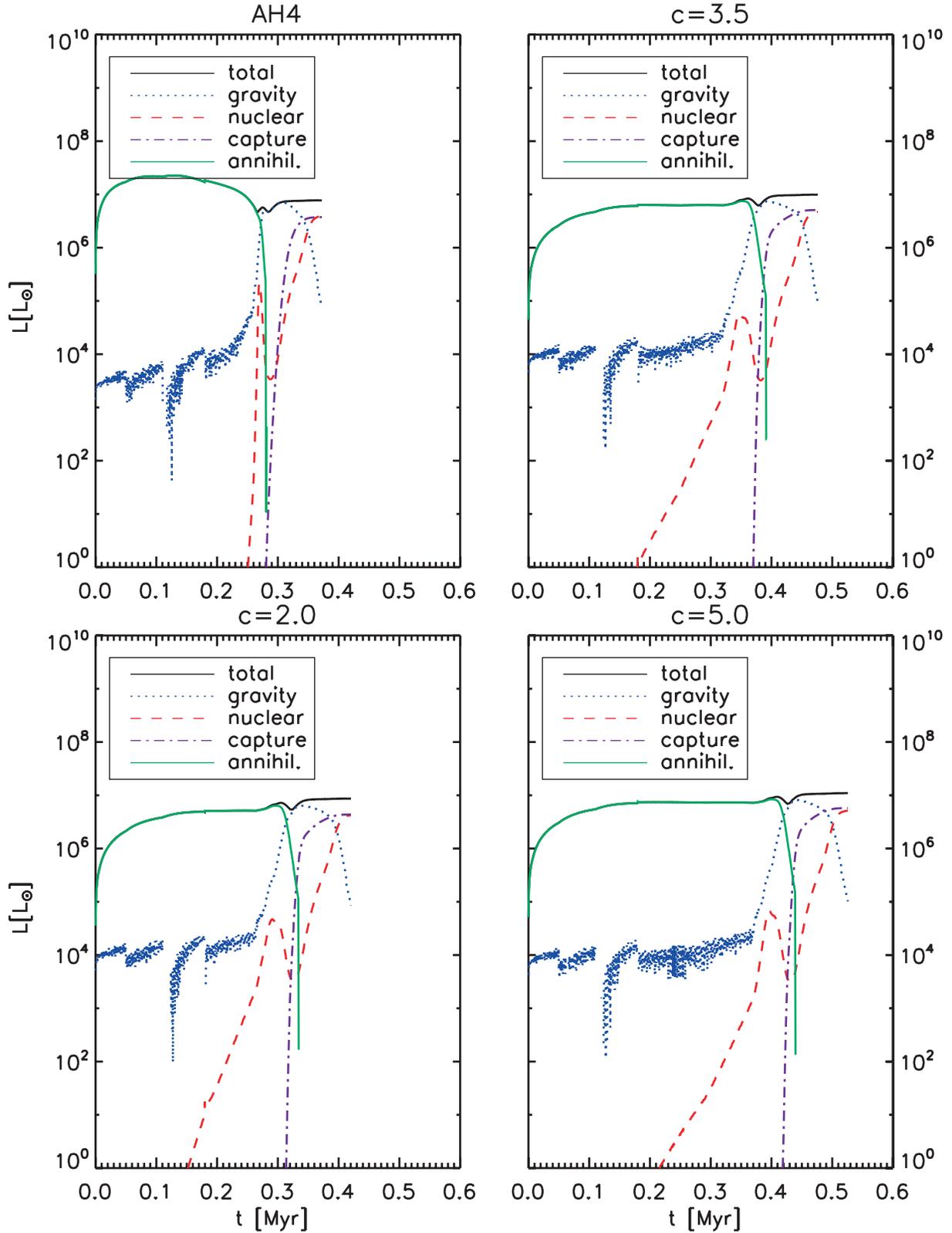}
\caption{Luminosity evolution as a function of time for the four cases under consideration. 
 The upper left panel displays the only boosted case, the AH4 model, with  $B=1500$, $m_{\chi}=2.35$ TeV and $c=3.5$.
 The other panels correspond to the unboosted case of a  $100$ GeV WIMP with $\sv=\svcanonical$ 
 and concentration parameters as labeled.}
\label{evol}
\end{figure}

In Fig. \ref{evol} we compare the luminosity evolution for the four cases we have just described. In all cases, DM annihilation heating provides the dominant contribution to the 
DS luminosity until the DM runs out. At this point Kelvin-Helmholtz contraction sets in and 
briefly provides the dominant heat source, until the star becomes hot and dense enough for fusion to begin.  In the cases where we include capture, DM annihilation may again become important at the later stages leading to a new DS phase.

In both upper panels  of Fig. \ref{evol} we take $c=3.5$: the left panel is the boosted AH4 model while the right panel is unboosted 
(the relative ratio of boost factors is effectively $64$ as mentioned before). Due to the ambiguity in the value of the concentration parameter we will also study its implications on  the evolution of the Dark Star  by running the unboosted $100$ GeV case for $c=(2, 3.5, 5)$ respectively; the lower two panels are the unboosted case with $c=2$ and 5.  The lower left panel  is the same as the canonical case studied in ~\citet{DSnl}.

{\it{Effects of Boost:}} The DM heating is the most powerful in the boosted AH4 case, since Eq.(1) indicates that heating scales with cross 
section.  Thus at any given time during the DS phase, this model has the brightest luminosity, as can be seen in Fig. \ref{evol}. 
 Consequently, the DM is burned up the more quickly in the boosted AH4 case, leading to the shortest DS lifetime.  
  
In order to balance the higher DM heating, a larger radius is required, which leads  to a lower central temperature and density.  See Fig.~\ref{dmdens}. 
As the adiabatically contracted DM runs out the evolution is  similar in all cases, the final luminosity being on the order of $10^7\lsun$.

 {\it{Effects of concentration parameter:}} When we compare the three different concentration parameter cases, we notice that the nuclear fusion sets in earlier for smaller values of $c$. The higher the concentration parameter the more adiabatically contracted DM available, therefore it takes longer to start the transition to the main sequence. This will lead to slightly higher final masses, which in turn translate to higher final luminosities. Nevertheless, as pointed out before they are all very close to $10^7\lsun$. In all three cases where $\sv=3\times 10^{-26}$cm$^3/$s we notice a flash in the luminosity when both the gravitational energy and the energy due to annihilations of the adiabatically contracted DM are relevant. This happens at a time somewhere between 0.31 Myr (for $c=2$ ) and 0.42 Myr (for $c=5$). The same flash can be seen for the AH4 panel, but now the true maximum of the luminosity
 occurs after only 0.1 Myr.  The maximum luminosity is $\sim 4\times 10^9 \lsun$.

The main differences between the different WIMP models as regards the luminosity evolution in
 Fig.\ref{evol} are in the time the ``pure'' dark star phase lasts. The higher the boost factor, the shorter this phase. Conversely  a larger concentration parameter $c$ prolongs the DS phase, since more DM is available. 

\subsubsection{Pre main sequence evolution, HR diagram}
\begin{figure}
\includegraphics[width=0.98\textwidth]{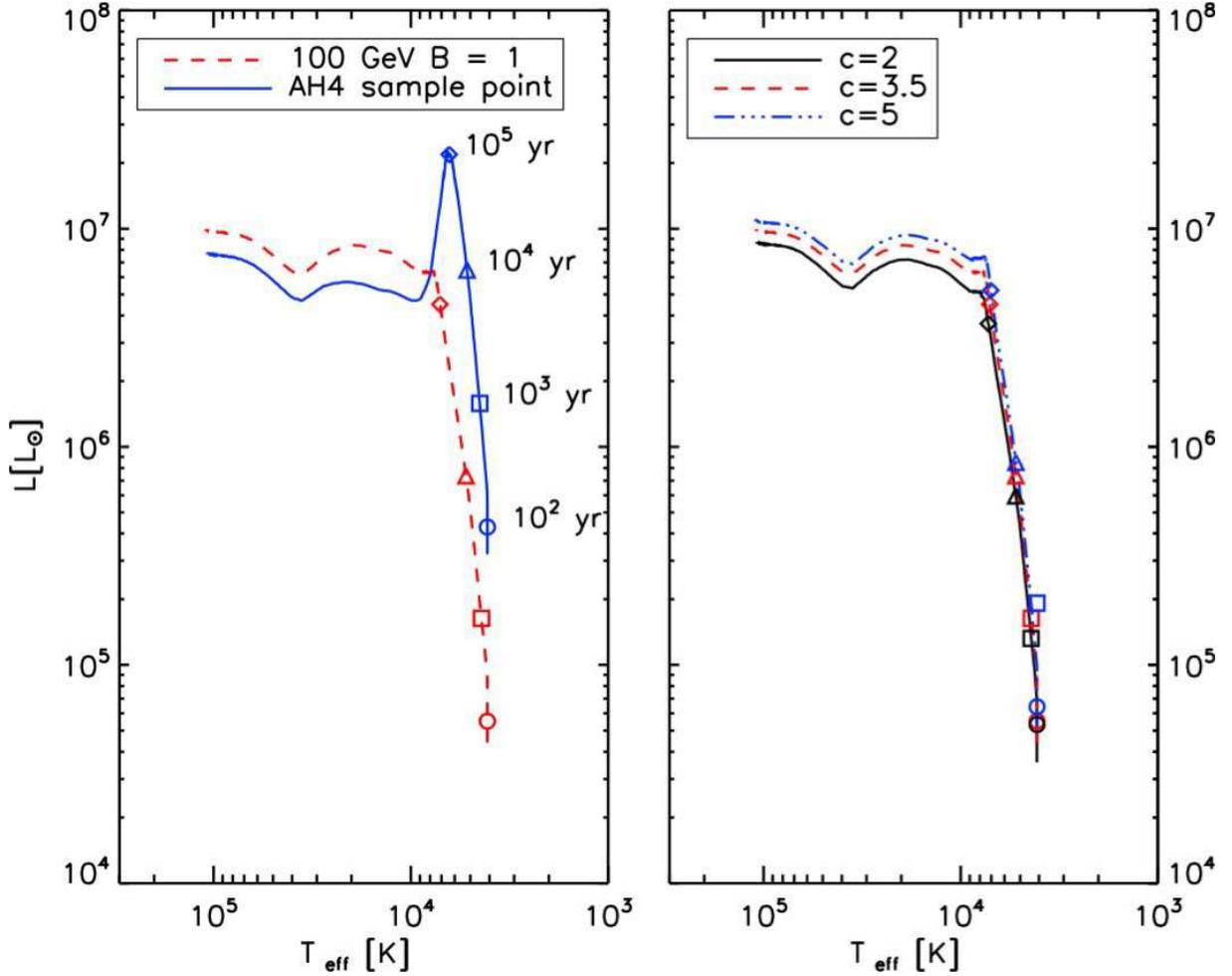}
\caption{Hertzsprung-Russell diagram for DS. The left panel displays the unboosted $100$ GeV case (dashed line) and the AH4 case,
both for c=3.5.  The right panel displays the unboosted 100 GeV case for a variety of concentration parameters as labeled.
Also labeled are a series of points corresponding to the evolution of the DS towars the ZAMS at different times.}
\label{HR}
\end{figure}

In Fig. \ref{HR} we plot Hertzsprung-Russell (H-R) diagrams for the four cases.  One can see two distinct phases.
 First, the DS goes up the Hayashi track with a very steep increase of the luminosity yet  relatively cool surface temperature,
$T_{eff}\leq 10^4$ K. At the end of the Hayashi track the star  enters the Henyey track. This path corresponds to an almost constant luminosity while the temperature increases fast, mostly due to the KH contraction phase. As a rule of thumb once a star is on the Heyney track its core should be fully radiative. The graphs end  at a temperature of $\sim 10^5$ K when  the star reaches the main sequence.

{\it{Effects of Boost:}} From the left panel of Fig. \ref{HR} one can see that the boosted AH4 case has the highest luminosity, due
to the extremely efficient DM heating $Q_{DM}\sim \sv/m_{\chi}$ forming a luminosity peak. 
However, as the AH4 case burns up its DM, the its luminosity falls.  The boosted and unboosted cases eventually cross over
at a temperature of $\sim 10^4$ K, and henceforth the unboosted case has a higher luminosity. Consequently, the boosted AH4 case has
the lowest luminosity as the star moves onto the main sequence, as discussed above.

{\it{Effects of concentration parameter:}} In the right panel of Fig. \ref{HR} the trend is uniform:  an increase in the concentration parameter leads to an increase in the luminosity. The difference is relatively small in the early stages of the evolution, at low temperatures. This is due to the fact that the adiabatically contracted DM density profile is not very sensitive to the concentration parameter, therefore about the same amount of DM heating will be generated in each case. However, for a lower value of the concentration parameter, the adiabatically contracted DM runs out faster, as there is less DM available. This leads to a shorter ``pure'' DS phase, as can also be seen from Fig. \ref{evol}, and consequently to slightly lower final mass and luminosities.

\subsubsection{Radius and Effective Temperature of DS}

\begin{figure}
\includegraphics[width=0.98\textwidth]{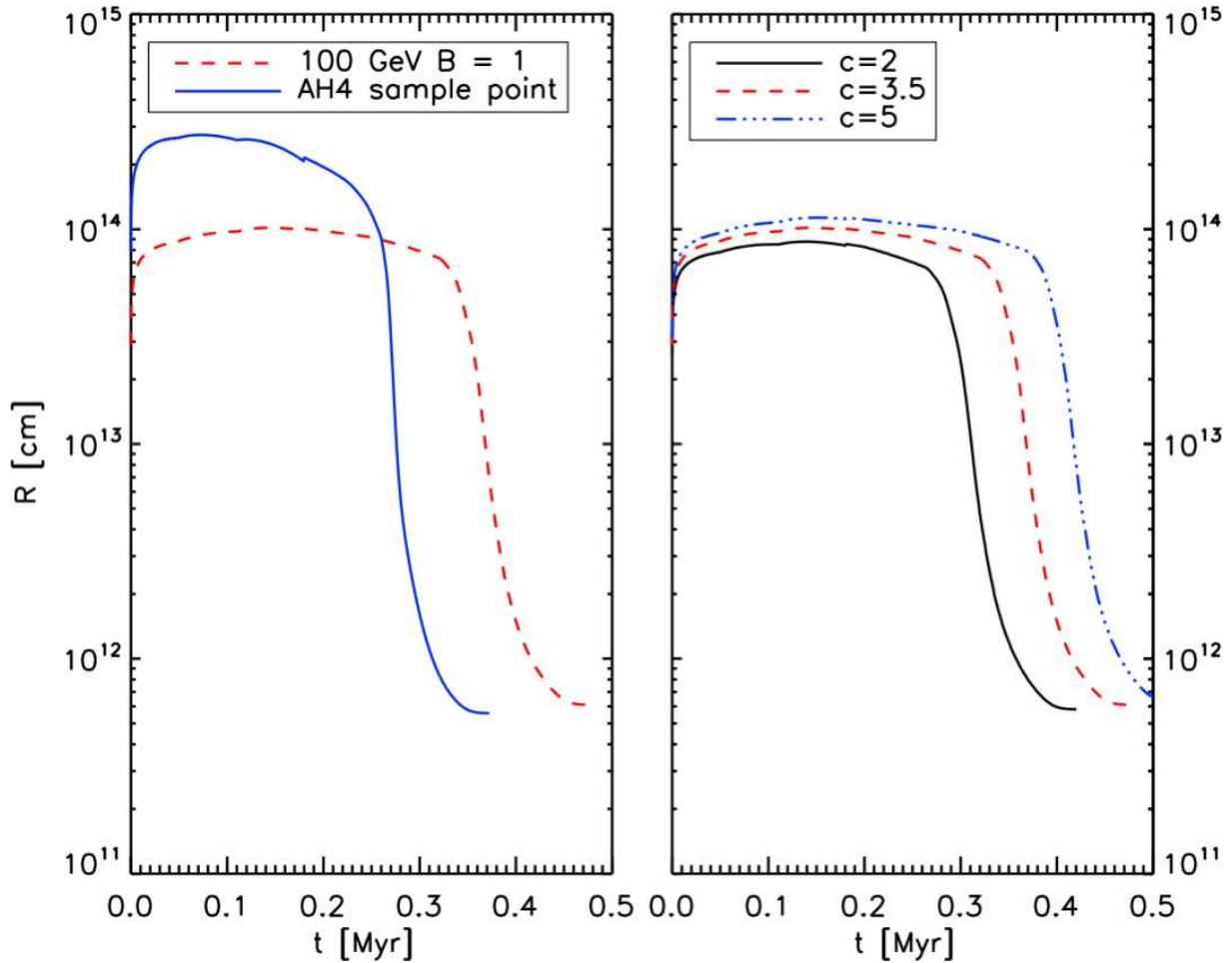}
\caption{DS radius as a function of time. The left panel displays the unboosted $100$ GeV  case (dashed line) 
and the AH4 case (solid line), both for $c=3.5$ The right panel corresponds to various concentration parameters $c$,  for the unboosted
 $100$ GeV  WIMP.}
\label{radius}
\end{figure}

{\it{Effects of Boost on Radii:}} In the ``pure'' DS phase the star will be puffier for a higher $\sv/m_{\chi}$ ratio, as can be seen from the left panel of Fig.\ref{radius}. This is expected due to the much higher DM heating in that case:  a larger radius is required to balance the DM energy
production and the radiated luminosity, which scales as $R^2$.
For instance, in the boosted AH4 case the maximum radius is at about $2\times 10^{14}$cm, whereas for the $100$ GeV non-boosted WIMP, the DS will have a maximum radius of $10^{14}$cm. However, as mentioned before, the KH contraction will set in earlier in the boosted case. This phase corresponds to the sharp decrease in radius in Fig.\ref{radius}. The final radii, as the DS enters the ZAMS are similar in both cases, at around $6\times 10^{11}$cm.  

{\it{Effects of Concentration Parameter on Radii:}} From the right panel of Fig.\ref{radius} we notice an uniform increase in the radius with the concentration parameter. Again, more efficient DM heating leads to a puffier DS. The maximum radii range from $8\times 10^{13}$cm to $10^{14}$cm for $c=(2-5)$. After the KH contraction phase the DS settles to a radius very close to $6\times 10^{11}$cm as it enters the ZAMS. 

\begin{figure}
\includegraphics[width=0.98\textwidth]{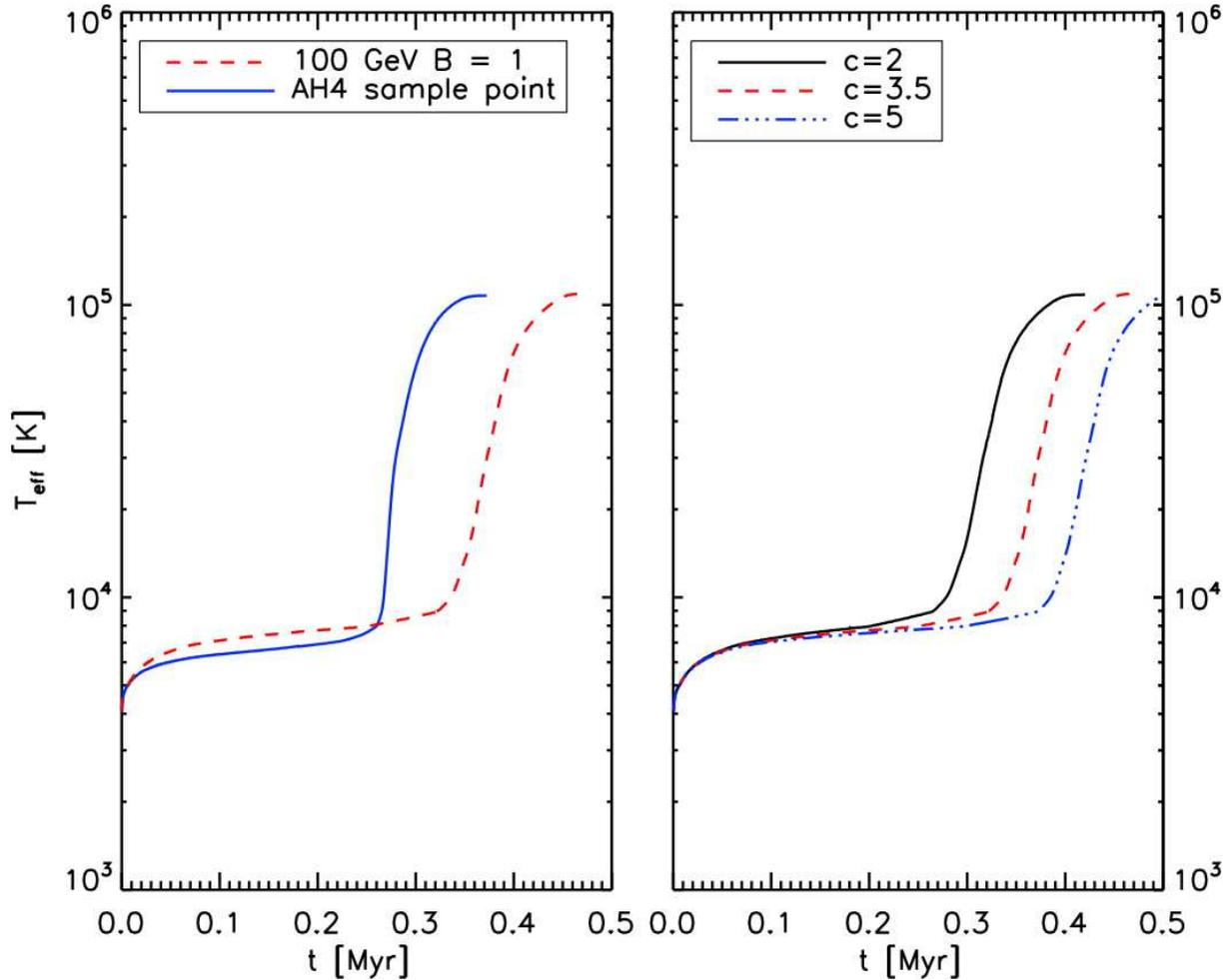}
\caption{Effective temperature as a function of time. The left panel corresponds to the unboosted 
$100$ GeV case (dashed line) and the boosted AH4 WIMP parameters defined previously (the solid line),
both for $c=3.5$. The right panel corresponds to various concentration parameters $c$,  for the unboosted
$100$ GeV WIMP.}
\label{Teff}
\end{figure}

{\it{Effective Temperatures:}} In Fig. \ref{Teff} we plot the effective surface temperature as a function of time. 
At first, during the DS phase, $T_{eff}$ is relatively constant below $10^4$K. Once the DM starts to run out, the star contracts and heats
up, leading to a sharp increase in $T_{eff}$ due to the onset of the KH contraction phase. Once nuclear fusion becomes the dominant energy supply and the star ceases to contract, the surface temperature reaches a plateau. The final value of  $T_{eff} \sim 10^5$K is always the
same for all cases, regardless of value of boost factor or concentration parameter. The AH4 case (left panel in Fig.\ref{Teff}) starts as a cooler star, again typical of more efficient DM heating in that case due to the boost factor. In addition, in the AH4 case the DM runs out more quickly, leading
 to an earlier increase in temperature. Regarding the various concentration parameters: during the DS phase, the surface temperature is roughly
 the same in  all cases,  $\sim 7\times10^3 $K.  The DS phase lasts the longest for the highest value of $c$ as this case has the most DM
 to begin with; thus the temperature starts to rise later for ever larger concentration parameters $c$.

\subsubsection{Energy Transport near the Core}

\begin{figure}
\begin{center}
\includegraphics[width=0.58\textwidth]{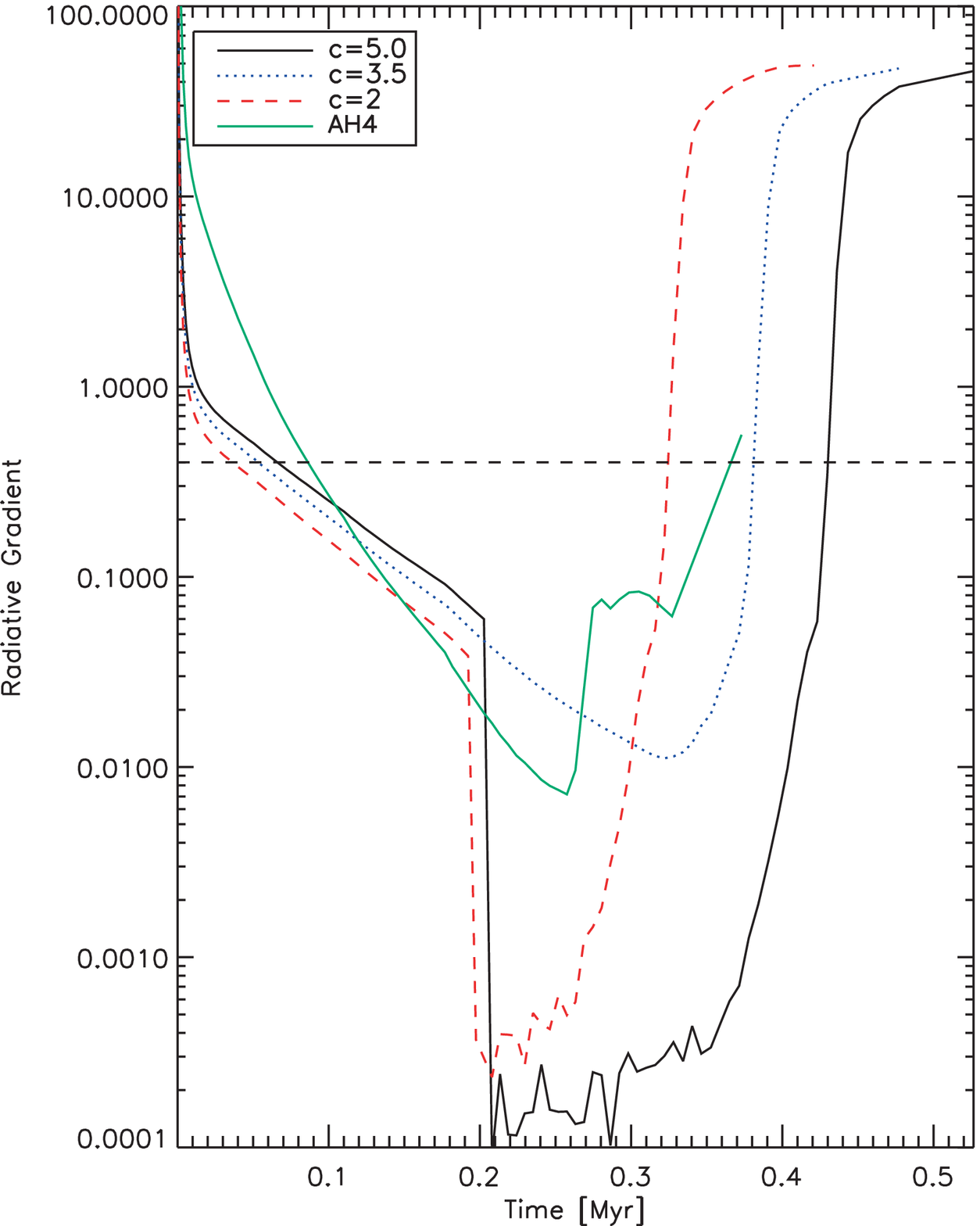}
\caption{Radiative gradient at the center of the DS as a function of time. 
Models with the gradient above the horizontal line are unstable to convection. 
The lines labeled by the value of the concentration parameter $c$ correspond to a DS powered by  $100$ GeV WIMPs with
canonical unboosted cross section. The model AH4 is  as defined previously in Fig. 1.\label{Fig_radgrad}}
\end{center}
\end{figure}

The DS starts with a fully convective structure, modeled by a fluid  with a polytropic index $n=1.5$. Then  a radiative core starts to develop, that grows outwards, until most of the star is described by a polytrope of index $n=3$. In Fig. \ref{Fig_radgrad} we plot the radiative gradient in the innermost zone at the center of the DS. The dashed horizontal line illustrates the critical value for 
convection.  Models above the line have a convective core, while models below the line have radiative cores. 
Models with more efficient DM heating -- i.e. the models with higher values of $c$ or the AH4 model --- 
transition to radiative energy transport later;  stars with more efficient DM heating require a larger radius. At a fixed luminosity with more efficient DM heating,  the star  must have a larger radius and a lower densities to keep DM heating and the stellar luminosity in balance.  With a lower density, the star has a lower central temperatures. At lower temperatures, the number of bound states increases, which increases the number of bound-free transitions.  Also, the number of free-free transitions increase.  These two effects increase the opacity which produces a larger radiative gradient and delays the transition from convective to radiative transport.

Once the original DM in the star runs out and nuclear fusion begins, a convective core develops in all cases.  The central gradient
is high due to the fact that nuclear fusion takes place primarily in the core of the star. Similarly, once the star repopulates its DM due 
to capture, this new DM population is thermalized with the star and its density is also sharply peaked at the center of the star. 
Thus both nuclear fusion and captured DM lead to a large radiative gradient in the core and therefore favor convection.

Again, the transition happens later for higher values of $c$.  
The time when the convective core develops corresponds almost precisely to the time when the captured DM heating becomes significant, as can be seen by comparing with Fig. \ref{evol}:   $0.31$ Myr, $0.37$ Myr and $0.42$ Myr for $c=2, 3.5$ and $5$ respectively. 

\subsubsection{Baryonic Central Density}
\begin{figure}
\includegraphics[width=0.98\textwidth]{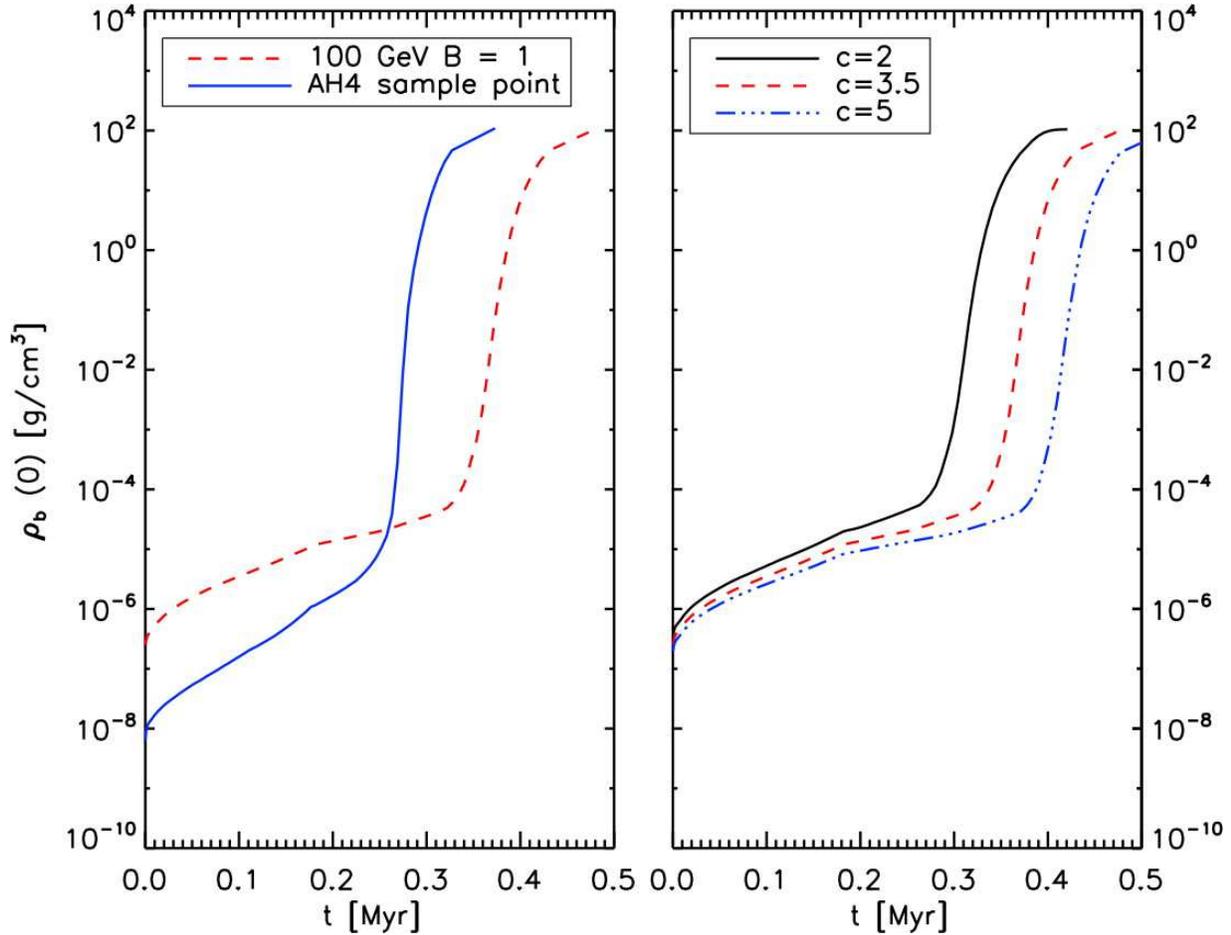}
\caption{Central baryon density as a function of time.   The left panel displays the unboosted $100$ GeV WIMP case (dashed line) and the boosted AH4 case (the solid line), both for c=3.5. 
The right panel corresponds to  various concentration parameters $c$ for the $100$ GeV WIMP case with unboosted canonical cross section.}
\label{bcd}
\end{figure}

The baryon central density is plotted in Fig. \ref{bcd} for the four cases we considered. 

The higher cross section of the AH4 case at first leads to a puffier star (larger radius to keep the radiated luminosity at a level to balance the
higher DM heating); thus it is not surprising that the AH4 case initially has a relatively lower central density.  However,  the initial DM
in the DS runs out earlier in the AH4 case due to more efficient burning; hence in the left panel the two lines cross
due to the earlier onset of the KH contraction (marked in the plot by the sharp increase of the densities) in the AH4 case.

The central density $\rho_b(0)$ scales inversely with the concentration parameter, as can be seen from the right panel 
  of the same plot. Again, more DM heating (higher $c$) will lead to larger radii, therefore smaller densities. However, as opposed to the situation depicted in the left panel, 
 In the right panel, the curves do not cross since the models with a larger concentration parameter have more DM, which delays the onset Helmholtz contraction. We will come back to this in Sec.\ref{SecDensProf}. In all cases considered here once the star goes onto the main sequence, the central density is close to 100g/cm$^3$.

\subsubsection{Mass as DS enters the ZAMS}

\begin{figure}
\includegraphics[width=0.98\textwidth]{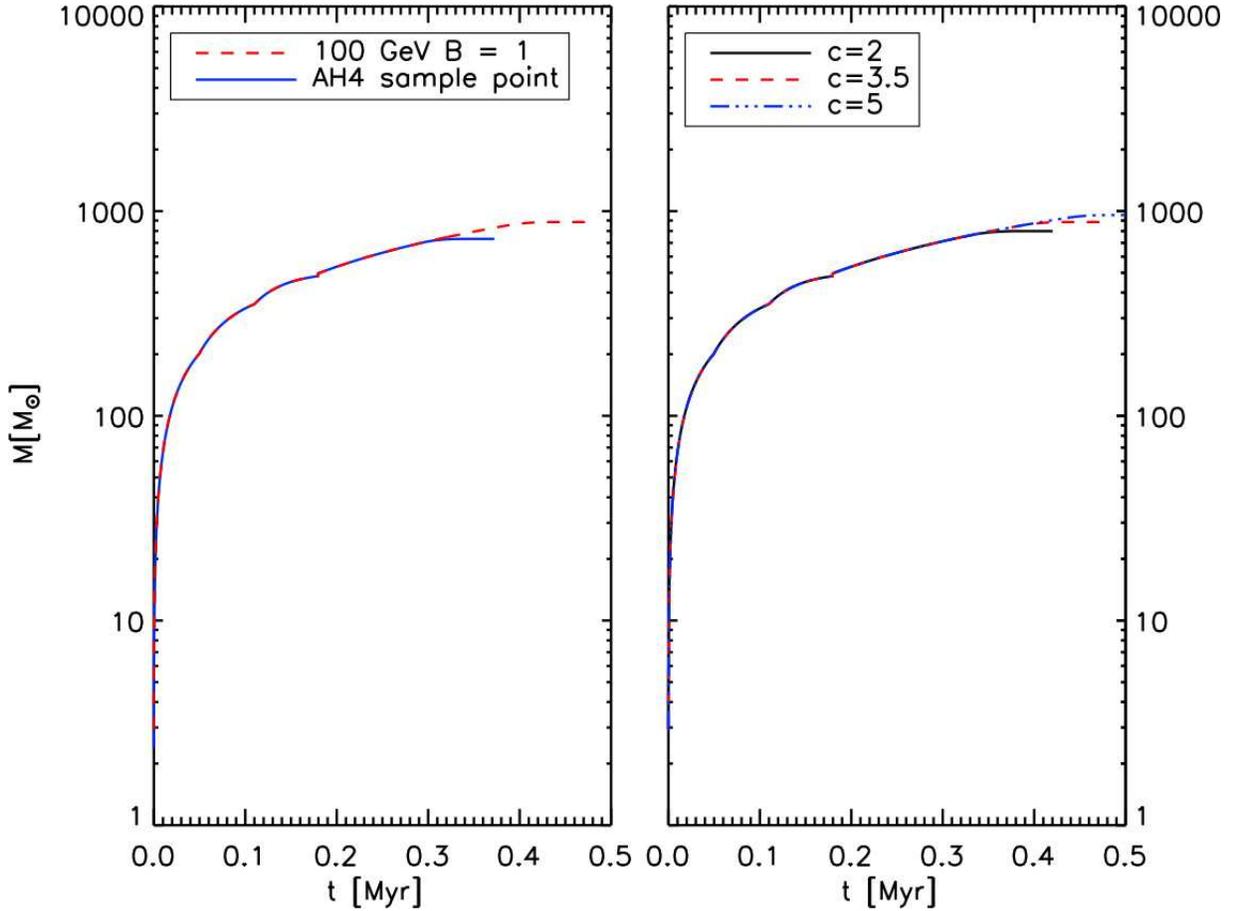}
\caption{DS mass as a function of time. The left panel displays the unboosted $100$ GeV  case (dashed line) and the AH4 model,
 both for c=3.5.  The right panel displays the unboosted $100$ GeV case for a variety of concentration parameters $c$.}
\label{accretion}
\end{figure}

In Fig.~\ref{accretion}, we have plotted the DS mass as a function of time.  In all cases, the final mass when the DS enters the main sequence
is $\sim 700-1000 \msun$; however, there are slight differences for different models.  The models with more
effective DM heating -- i.e. the AH4 model compared to the $100$ GeV unboosted case -- burn up their original adiabatically contracted
DM the most quickly and enter the KH contraction phase the soonest. This in turn leads to a smaller final mass, as feedback effects will shut off accretion sooner. 
When comparing the cases with different concentration parameter we notice that the DS final mass is an increasing
 function of $c$. As previously explained an increase in the concentration parameter leads to a longer DS phase, and hence to more mass accreted. 
For the case of the unboosted $100$ GeV WIMP with $c=3.5$ the final mass is around $900 \msun$, whereas for the $AH4$ case it
 is close to $700 \msun$. For $c$ ranging from $2-5$ the DS will have a mass in the $800\msun-1000\msun$ as it reaches the main sequence.
  
\subsubsection{Density Profiles for DM and Baryons inside the DS; amount of adiabatically contracted DM}\label{SecDensProf}
\begin{figure}
\includegraphics[width=0.98\textwidth]{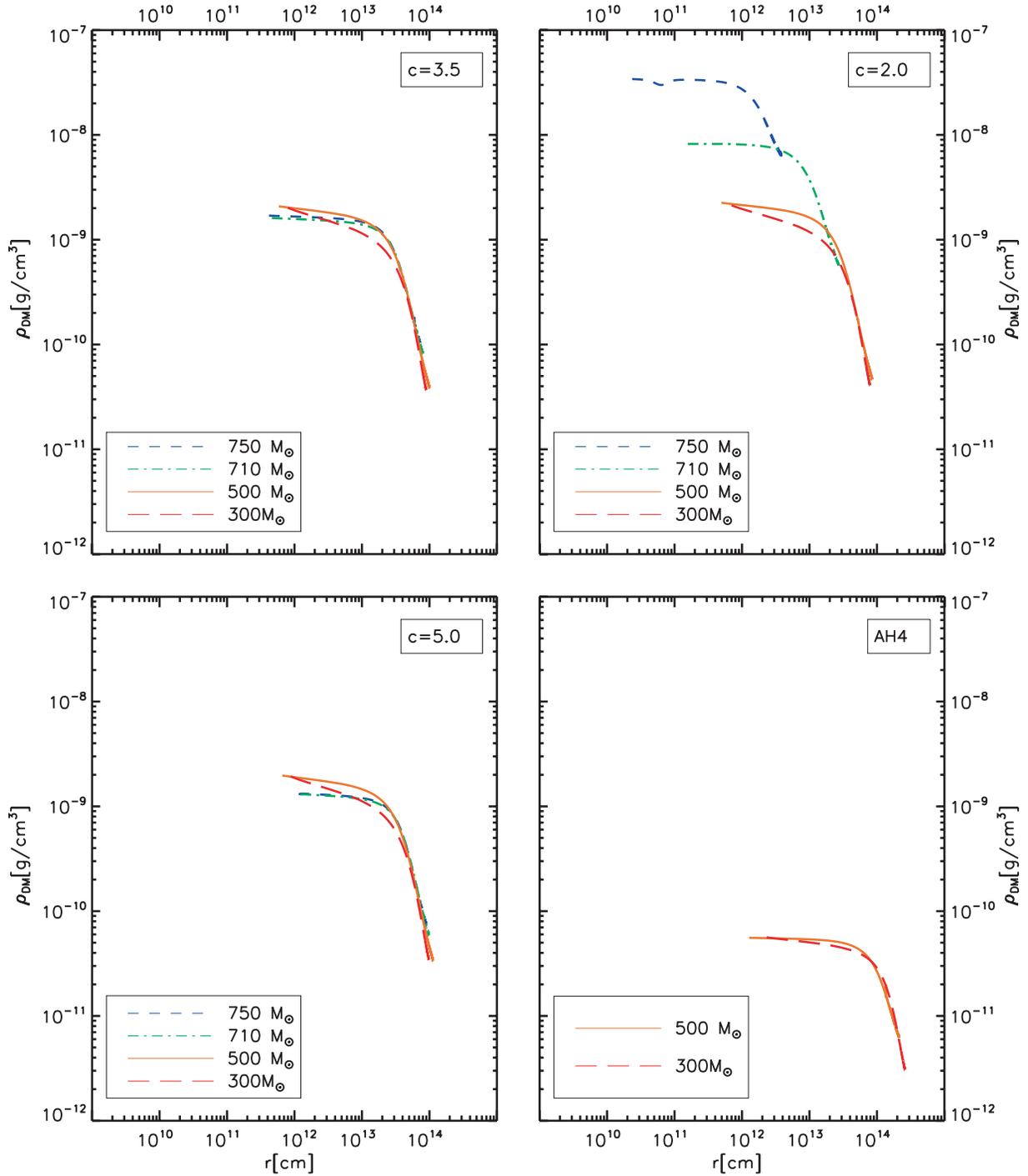}
\caption{Adiabatically contracted DM density profiles. Each line corresponds to a fixed value of the mass of the DS during its evolution. Note that certain lines that are mentioned in the legend do not appear plotted in all four panels. This is due to the fact that at that stage the DS has exhausted all the DM.}
\label{dmdens}
\end{figure}

\begin{figure}
\includegraphics[width=0.98\textwidth]{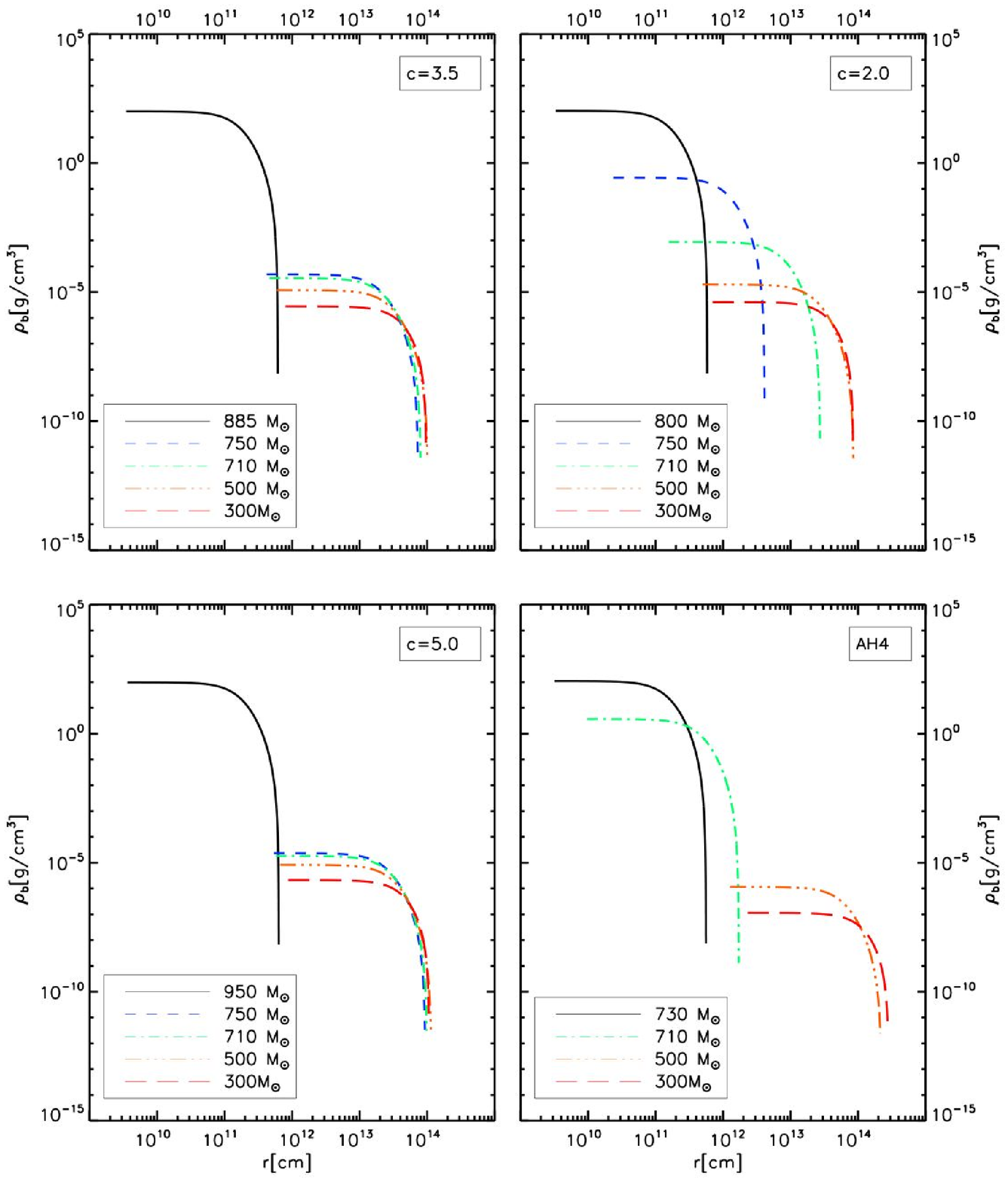}
\caption{Baryonic density profiles at different stellar masses for the four cases under consideration. The solid line corresponds to the mass of the DS as it enters the ZAMS.}
\label{dens}
\end{figure}

In  Fig. \ref{dmdens} and Fig.\ref{dens}, we have plotted the density profiles of the  adiabatically contracted DM and baryons respectively.  
The AH4 model  for the same stellar mass has a lower DM density than the canonical unboosted $100$ GeV case by
 roughly an order of magnitude, and also has a more extended profile.  For instance, in the case of a DS of $300 \msun$ the
  values are $5\times 10^{-11}$g/cm$^3$ (AH4) and $\sim 1\times 10^{-9}$g/cm$^3$ (canonical) respectively. This effect can be 
  attributed to the fact that at a higher annihilation cross sections or a lower particle masses\footnote{
A similar effect was noticed in ~\citet{DSnl}, where it was found that ``the average DM density in the star is an increasing function of $M_{\chi}$''; n.b. the higher annihilation cross section of the AH4 case can be traded for a lower WIMP mass since heating scales as
$\langle \sigma v \rangle /M_\chi$. }, a larger radius and a lower density are needed to balance the DM heating and the stellar luminosity (which scales as $R^2$).

During the early stages of the DS evolution the dependence of the adiabatically contracted DM density on the concentration parameter  is very small at least for the range we have considered here. As it can be seen from Fig.\ref{bcd}, prior to the onset of the KH contraction phase,
 the central baryon densities for models with different values of the concentration parameter   have similar baryon density and DM
  density as well.  Models with a larger concentration parameter have slightly more dark matter, have slightly lower central DM densities, 
  and are also more extended.  Before the contraction phase, the central DM density of the $c=5$ case's density is 10\% lower than the $c=2$ case. 
   The radius is 20\% larger.  Models with different concentration parameters only begin to dramatically diverge once the star begins to contract and enters its Kelvin-Helmholz contraction phase. At this point, the star begins to shrink, which cause the DM densities to increase dramatically. 

  DS in halos with a larger concentration parameter have more DM and thus delay the onset of the KH phase. For $c=2$, the contraction phase begins  at $t\sim 0.28$ Myr;  See Fig.\ref{radius}.  At this time, the stellar mass has reached $\sim 700\msun$. For $c=3.5$ and 5,  the contraction phase begins later.  In the case of $c=5$, the star has a mass of $\sim850\msun$.  Thus the contraction begins once the star is more massive.   

At a fixed stellar mass, the DM densities  will differ dramatically between models which are in the contraction phase compared to 
 those which are not contacting .  For instance, let us consider a $750 \msun$ DS.  In the case of $c=2$,
the star has entered the contraction phase.  While the cases with a larger concentration parameter (3.5,5)  have
 yet to begin their contraction phase.  Hence the stars with a larger concentration, have a lower DM density and are more extended, which can be seen in Fig.~\ref{dmdens}.

\begin{figure}
\includegraphics[width=0.98\textwidth]{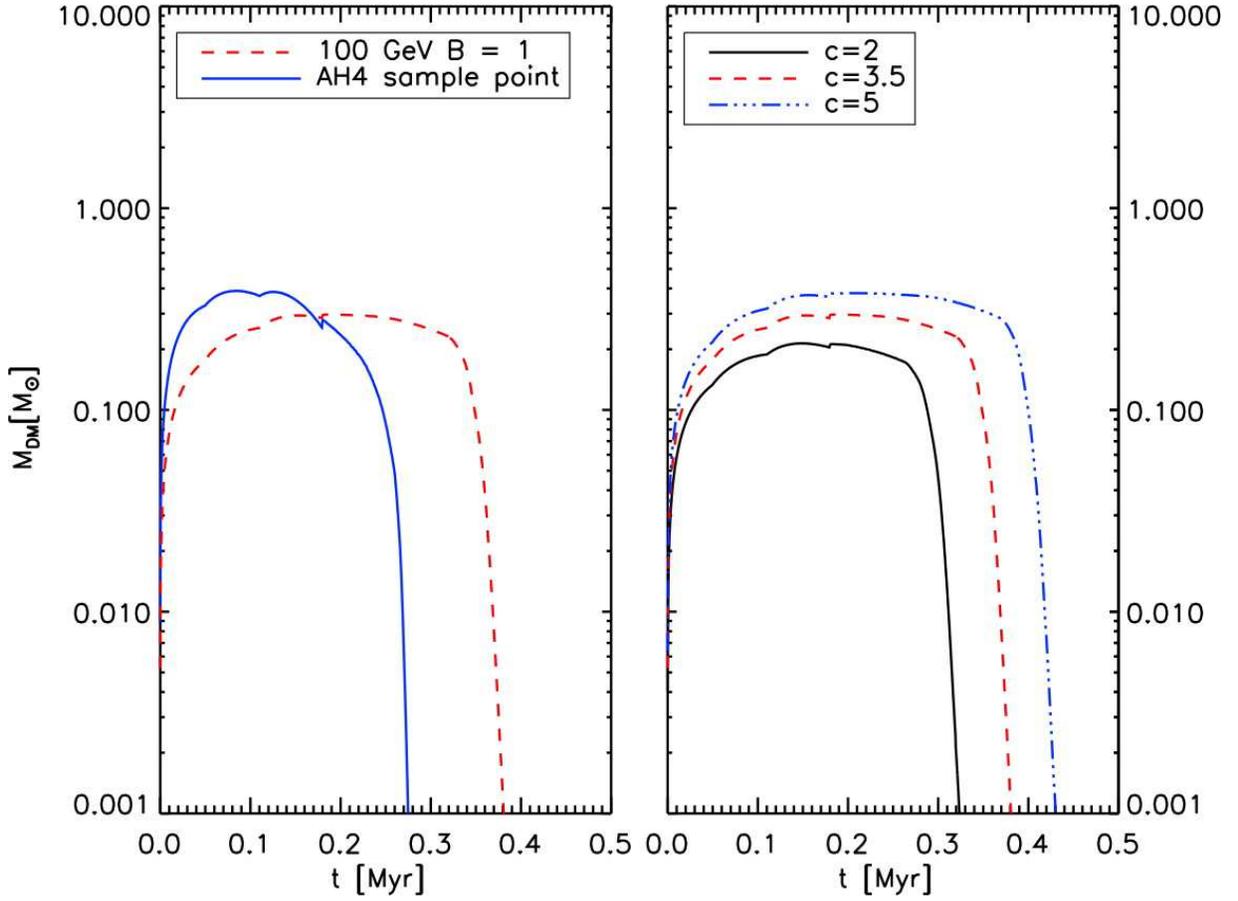}
\caption{Amount of adiabatically contracted DM inside the star as a function of time.  The left panel displays the unboosted $100$ GeV 
 case (dashed line) and the AH4 model (solid line), both for $c=3.5$ The right panel displays the unboosted 100 GeV
 case for a variety of concentration parameters $c$.}
 \label{DM}
\end{figure}

Finally, in Fig. \ref{DM} we have plotted the amount of adiabatically contracted DM inside the DS as a function of time.	 
 One can also see that DM densities are many orders of magnitude lower than their baryonic counterparts at all times.  
 Although the amount of DM never exceeds  $0.4 \msun$, yet this is sufficient to power the DS all the way up to $\sim 1000 \msun$ (where most of the
mass is baryons) and $10^7 L_\odot$.  Indeed DM heating is a very powerful energy source.

\clearpage
\section{Summary and Conclusions}
\label{sec:summary}
In this paper we have considered the effect on DS of leptophilic boosted DM annihilation cross sections, as would typically be required to explain
PAMELA data.  Second, we have varied the concentration parameter in a host of DS models.  We have restricted our study to include 
the following two sources of DM: i) the DM originally contained in the star due to adiabatic contraction and ii) the minimal capture scenario.
We have not included the additional DM due to extended adiabatic contraction or to maximal capture models discussed in ~\citet{SMDS}.
Nonetheless, the dependence of DS properties on boost factor and concentration parameter can easily be seen in this paper.  As our 
prototypical boosted case, we have focused on the AH4 model with the following parameters:  $B=1500$, $m_{\chi}=2.35$ TeV and $c=3.5$.
In the unboosted cases, we have taken $M_\chi = 100$ GeV and three values of the concentration parameter, $c = 2, 3.5,$ and $5$.

We have found that, if the positron excess observed by PAMELA is indeed due to leptophilic boosted DM, then there would be
an early DS phase of stellar evolution powered by DM heating, lasting long enough to bring the star to substantially higher mass and luminosity
 than predicted for regular Pop.III zero metallicity stars. Our basic results are that the final stellar properties, after the DS runs out of 
 or its original adiabatically contracted DM fuel, undergoes Kelvin Helmholtz contraction and enters the main sequence,
 are always roughly the same: $\sim 1000 \msun$, $\sim 10^7 L_\odot$, lifetime $\sim 10^6$ yrs.  Similarly, these same DS properties are also
the basic result independent of the value of the concentration parameter in the range between c=2 and c=5.

We reiterate that these values are only for the case of simple adiabatic contraction and minimal capture; if we were to include the
additional DM due to extended AC or maximal capture, then these values would be  different by many orders
of magnitude. However, the basic dependence on the parameters of interest in this paper would generalize. In particular, the result
that the final mass, luminosity, and lifetime are relatively {\it independent} of boosted cross section or value of $c$ would still hold up. In 
addition, the slight differences from model to model, discussed in the next paragraph, would also hold up.

We have found that the lifetime, final mass, and final luminosity  of the DS, though roughly similar in all cases, do show some dependence on 
boost factor and concentration parameter.  We have found that the DS lifetime is shorter in the boosted case, 
since the DM is exhausted sooner. On the other hand, the lifetime is longer for higher concentration parameter since there is more 
DM to begin with. Thus nuclear burning starts earlier in the case of unboosted cross section or low concentration parameter.   
The DS accretes matter continuously while
it remains powered by DM heating. Hence the largest final stellar mass results for the longest living DS, i.e. the unboosted case with the
highest values of $c$. In all cases, the final mass is $\sim 1000 \msun$.  We have shown the H-R diagram for all cases, studying 
both the Hayashi track and the approach to fusion power.
We have showed that the `final' luminosity at the end of the DS lifetime is $\sim 10^7 L_\odot$ in all cases, with the luminosity decreasing
slightly as a function of increasing boost factor or decreasing $c$, again because of the more rapidly depleted pool of DM.
We have also examined the luminosity evolution of the DS as a function of time.  During the DS phase itself, the luminosity is higher/lower
for boosted/unboosted cross sections. The reduced luminosity during the DS evolution
 in the unboosted case  is a consequence of the reduced energy production,
since DM heating is proportional to annihilation cross section and the square of the DM density.  
Then at lower cross section (unboosted), a smaller radius is needed
to balance the DM energy production and the radiated luminosity.  Consequently the unboosted case 
has higher central densities (both
for baryons and DM).  Similarly, lower values of $c$ have lower luminosity during the DS phase, smaller radii, and higher central density.
In all cases the DM density is a minute fraction of the total density, with baryons dominating the gravitational potential;
we have shown the density profiles of both components.  Again in all cases, the amount of DM inside the star never amounts to more
than 0.4$\msun$, a tiny fraction of a star that grows to $\sim 1000 \msun$; yet this DM is sufficient to power the star. This is "the power of darkness."

\section*{Acknowledgements}
We acknowledge support from: the DOE and the Michigan Center for Theoretical Physics at the University of MI (C.I. and K.F.); DOE at Fermilab (D.S.).  We thank Pearl  Sandick for helpful conversations. 

\bibliographystyle{apj}  
\bibliography{RefsDS}   

\begin{thebibliography}{122}
\expandafter\ifx\csname natexlab\endcsname\relax\def\natexlab#1{#1}\fi

\bibitem[{Abazajian {et~al.}(2010)Abazajian, Agrawal, Chacko, \&
  Kilic}]{Abazajian2010}
Abazajian, K.~N., Agrawal, P., Chacko, Z., \& Kilic, C. 2010, arXiv:1002.3820

\bibitem[{Abdo {et~al.}(2009{\natexlab{a}})}]{Abdo2009}
Abdo, A.~A., {et~al.} 2009{\natexlab{a}}, Phys. Rev. Lett., 103, 251101

\bibitem[{Abdo {et~al.}(2009{\natexlab{b}})}]{FERMI}
---. 2009{\natexlab{b}}, Phys. Rev. Lett., 102, 181101

\bibitem[{Abdo {et~al.}(2010)}]{Abdo2010}
---. 2010, JCAP, 1004, 014

\bibitem[{Adriani {et~al.}(2009{\natexlab{a}})}]{PAMELApap}
Adriani, O., {et~al.} 2009{\natexlab{a}}, Phys. Rev. Lett., 102, 051101

\bibitem[{Adriani {et~al.}(2009{\natexlab{b}})}]{PAMELA}
---. 2009{\natexlab{b}}, Nature, 458, 607

\bibitem[{Adriani {et~al.}(2010)}]{Adriani2010}
---. 2010, arXiv:1001.3522

\bibitem[{Ahlers {et~al.}(2009)Ahlers, Mertsch, \& Sarkar}]{Ahlers2009}
Ahlers, M., Mertsch, P., \& Sarkar, S. 2009, Phys. Rev., D80, 123017

\bibitem[{Arkani-Hamed {et~al.}(2009)Arkani-Hamed, Finkbeiner, Slatyer, \&
  Weiner}]{AHDM}
Arkani-Hamed, N., Finkbeiner, D.~P., Slatyer, T.~R., \& Weiner, N. 2009, Phys.
  Rev., D79, 015014

\bibitem[{{Bahcall}(1989)}]{bahcall}
{Bahcall}, J.~N. 1989, {Neutrino astrophysics} (Cambridge and New York,
  Cambridge University Press, 1989, 584 p.)

\bibitem[{Bai {et~al.}(2009)Bai, Carena, \& Lykken}]{Bai2009}
Bai, Y., Carena, M., \& Lykken, J. 2009, Phys. Rev., D80, 055004

\bibitem[{Bajc {et~al.}(2010)Bajc, Enkhbat, Ghosh, Senjanovic, \&
  Zhang}]{Bajc2010}
Bajc, B., Enkhbat, T., Ghosh, D.~K., Senjanovic, G., \& Zhang, Y. 2010, JHEP,
  05, 048

\bibitem[{Baltz {et~al.}(2002)Baltz, Edsjo, Freese, \& Gondolo}]{Baltz2001}
Baltz, E.~A., Edsjo, J., Freese, K., \& Gondolo, P. 2002, Phys. Rev., D65,
  063511

\bibitem[{Barger {et~al.}(2009{\natexlab{a}})Barger, Gao, Keung, Marfatia, \&
  Shaughnessy}]{Barger2009}
Barger, V., Gao, Y., Keung, W.~Y., Marfatia, D., \& Shaughnessy, G.
  2009{\natexlab{a}}, Phys. Lett., B678, 283

\bibitem[{Barger {et~al.}(2009{\natexlab{b}})Barger, Keung, Marfatia, \&
  Shaughnessy}]{Ref:annih1}
Barger, V., Keung, W.~Y., Marfatia, D., \& Shaughnessy, G. 2009{\natexlab{b}},
  Phys. Lett., B672, 141

\bibitem[{Barkana \& Loeb(2001)}]{BarLoe01}
Barkana, R., \& Loeb, A. 2001, Phys. Rept., 349, 125

\bibitem[{{Barnes} \& {White}(1984)}]{blumenthal2}
{Barnes}, J., \& {White}, S.~D.~M. 1984, \mnras, 211, 753

\bibitem[{Barwick {et~al.}(1997)}]{HEAT}
Barwick, S.~W., {et~al.} 1997, Astrophys. J., 482, L191

\bibitem[{Bergstrom {et~al.}(2009)Bergstrom, Edsjo, \&
  Zaharijas}]{DMinterpdata}
Bergstrom, L., Edsjo, J., \& Zaharijas, G. 2009, Phys. Rev. Lett., 103, 031103

\bibitem[{Bertone \& Fairbairn(2008)}]{bertone}
Bertone, G., \& Fairbairn, M. 2008, Phys. Rev., D77, 043515

\bibitem[{Bi {et~al.}(2009)}]{Bi2009}
Bi, X.-J., {et~al.} 2009, Phys. Rev., D80, 103502

\bibitem[{Blasi(2009)}]{Blasi09}
Blasi, P. 2009, Phys. Rev. Lett., 103, 051104

\bibitem[{Blumenthal {et~al.}(1986)Blumenthal, Faber, Flores, \&
  Primack}]{blumenthal1}
Blumenthal, G.~R., Faber, S.~M., Flores, R., \& Primack, J.~R. 1986, Astrophys.
  J., 301, 27

\bibitem[{{Boulares}(1989)}]{boulares}
{Boulares}, A. 1989, Astrophys. J., 342, 807

\bibitem[{{Bouquet} \& {Salati}(1989)}]{bouquet}
{Bouquet}, A., \& {Salati}, P. 1989, \apj, 346, 284

\bibitem[{Bromm \& Larson(2004)}]{BroLar04}
Bromm, V., \& Larson, R.~B. 2004, Ann. Rev. Astron. Astrophys., 42, 79

\bibitem[{Catena {et~al.}(2009)Catena, Fornengo, Pato, Pieri, \&
  Masiero}]{Catena2009}
Catena, R., Fornengo, N., Pato, M., Pieri, L., \& Masiero, A. 2009,
  arXiv:0912.4421

\bibitem[{{Chandrasekhar}(1939)}]{chandra}
{Chandrasekhar}, S. 1939, {An introduction to the study of stellar structure}
  (Chicago, Ill., The University of Chicago press [1939])

\bibitem[{Chang {et~al.}(2008)}]{ATIC}
Chang, J., {et~al.} 2008, Nature, 456, 362

\bibitem[{Chen {et~al.}(2010)Chen, Mandal, \& Takahashi}]{Ref:ChenJCAP10}
Chen, C.-R., Mandal, S.~K., \& Takahashi, F. 2010, JCAP, 1001, 023

\bibitem[{Chen {et~al.}(2009{\natexlab{a}})Chen, Nojiri, Takahashi, \&
  Yanagida}]{decay1}
Chen, C.-R., Nojiri, M.~M., Takahashi, F., \& Yanagida, T.~T.
  2009{\natexlab{a}}, Prog. Theor. Phys., 122, 553

\bibitem[{Chen \& Takahashi(2009)}]{decay3}
Chen, C.-R., \& Takahashi, F. 2009, JCAP, 0902, 004

\bibitem[{Chen {et~al.}(2009{\natexlab{b}})Chen, Takahashi, \&
  Yanagida}]{decay2}
Chen, C.-R., Takahashi, F., \& Yanagida, T.~T. 2009{\natexlab{b}}, Phys. Lett.,
  B671, 71

\bibitem[{Cholis {et~al.}(2009{\natexlab{a}})Cholis, Dobler, Finkbeiner,
  Goodenough, \& Weiner}]{Ref:annih2}
Cholis, I., Dobler, G., Finkbeiner, D.~P., Goodenough, L., \& Weiner, N.
  2009{\natexlab{a}}, Phys. Rev., D80, 123518

\bibitem[{Cholis {et~al.}(2009{\natexlab{b}})Cholis, Finkbeiner, Goodenough, \&
  Weiner}]{Ref:annih5}
Cholis, I., Finkbeiner, D.~P., Goodenough, L., \& Weiner, N.
  2009{\natexlab{b}}, JCAP, 0912, 007

\bibitem[{Cholis {et~al.}(2009{\natexlab{c}})Cholis, Goodenough, Hooper, Simet,
  \& Weiner}]{Ref:annih8}
Cholis, I., Goodenough, L., Hooper, D., Simet, M., \& Weiner, N.
  2009{\natexlab{c}}, Phys. Rev., D80, 123511

\bibitem[{Cirelli {et~al.}(2008)Cirelli, Franceschini, \&
  Strumia}]{Cirelli2008}
Cirelli, M., Franceschini, R., \& Strumia, A. 2008, Nucl. Phys., B800, 204

\bibitem[{Cirelli {et~al.}(2009)Cirelli, Kadastik, Raidal, \&
  Strumia}]{Ref:annih4}
Cirelli, M., Kadastik, M., Raidal, M., \& Strumia, A. 2009, Nucl. Phys., B813,
  1

\bibitem[{Cirelli \& Strumia(2008)}]{Ref:annih3}
Cirelli, M., \& Strumia, A. 2008, arXiv:0808.3867

\bibitem[{{Clayton}(1968)}]{clayton}
{Clayton}, D.~D. 1968, {Principles of stellar evolution and nucleosynthesis}
  (New York: McGraw-Hill, 1968)

\bibitem[{de~Boer {et~al.}(2002)de~Boer, Sander, Horn, \& Kazakov}]{deBoer2002}
de~Boer, W., Sander, C., Horn, M., \& Kazakov, D. 2002, Nucl. Phys. Proc.
  Suppl., 113, 221

\bibitem[{Delahaye {et~al.}(2009)}]{delahayesalati}
Delahaye, T., {et~al.} 2009, arXiv:0905.2144

\bibitem[{Di~Bernardo {et~al.}(2009)Di~Bernardo, Gaggero, Grasso, \&
  collaboration}]{DiBernardo2009}
Di~Bernardo, G., Gaggero, D., Grasso, D., \& collaboration, f. t. F.-L. 2009,
  arXiv:0912.3887

\bibitem[{Diemand {et~al.}(2007)Diemand, Kuhlen, \&
  Madau}]{2007ApJ...667..859D}
Diemand, J., Kuhlen, M., \& Madau, P. 2007, Astrophys. J., 667, 859

\bibitem[{Diemand {et~al.}(2008)}]{Diemand2008}
Diemand, J., {et~al.} 2008, Nature, 454, 735

\bibitem[{Feldman {et~al.}(2009)Feldman, Liu, \& Nath}]{BWreso3}
Feldman, D., Liu, Z., \& Nath, P. 2009, Phys. Rev., D79, 063509

\bibitem[{Fox \& Poppitz(2009)}]{Ref:annih7}
Fox, P.~J., \& Poppitz, E. 2009, Phys. Rev., D79, 083528

\bibitem[{Freese {et~al.}(2008{\natexlab{a}})Freese, Bodenheimer, Spolyar, \&
  Gondolo}]{Freese:2008wh}
Freese, K., Bodenheimer, P., Spolyar, D., \& Gondolo, P. 2008{\natexlab{a}},
  Astrophys. J., 685, L101

\bibitem[{Freese {et~al.}(2009)Freese, Gondolo, Sellwood, \& Spolyar}]{DS3}
Freese, K., Gondolo, P., Sellwood, J.~A., \& Spolyar, D. 2009, Astrophys. J.,
  693, 1563

\bibitem[{Freese {et~al.}(2010)Freese, Ilie, Spolyar, Valluri, \&
  Bodenheimer}]{SMDS}
Freese, K., Ilie, C., Spolyar, D., Valluri, M., \& Bodenheimer, P. 2010,
  arXiv:1002.2233

\bibitem[{Freese {et~al.}(2008{\natexlab{b}})Freese, Spolyar, \&
  Aguirre}]{DMcap}
Freese, K., Spolyar, D., \& Aguirre, A. 2008{\natexlab{b}}, JCAP, 0811, 014

\bibitem[{Fujita {et~al.}(2009)}]{Fujita2009}
Fujita, Y., {et~al.} 2009, Phys. Rev., D80, 063003

\bibitem[{Gelmini \& Gondolo(2008)}]{Gelmini2008}
Gelmini, G.~B., \& Gondolo, P. 2008, JCAP, 0810, 002

\bibitem[{Gondolo {et~al.}(2010)Gondolo, Huh, Kim, \& Scopel}]{Gondolo2010}
Gondolo, P., Huh, J.-H., Kim, H.~D., \& Scopel, S. 2010, arXiv:1004.1258

\bibitem[{Grajek {et~al.}(2009)Grajek, Kane, Phalen, Pierce, \&
  Watson}]{Grajek2008}
Grajek, P., Kane, G., Phalen, D., Pierce, A., \& Watson, S. 2009, Phys. Rev.,
  D79, 043506

\bibitem[{Grasso {et~al.}(2009)}]{Grasso2009}
Grasso, D., {et~al.} 2009, Astropart. Phys., 32, 140

\bibitem[{Guo \& Wu(2009)}]{BWreso2}
Guo, W.-L., \& Wu, Y.-L. 2009, Phys. Rev., D79, 055012

\bibitem[{He(2009)}]{DMreview}
He, X.-G. 2009, Mod. Phys. Lett., A52, 2139

\bibitem[{Hisano {et~al.}(2004)Hisano, Matsumoto, \& Nojiri}]{Som3}
Hisano, J., Matsumoto, S., \& Nojiri, M.~M. 2004, Phys. Rev. Lett., 92, 031303

\bibitem[{Hooper {et~al.}(2009{\natexlab{a}})Hooper, Blasi, \&
  Serpico}]{Hooperetal08}
Hooper, D., Blasi, P., \& Serpico, P.~D. 2009{\natexlab{a}}, JCAP, 0901, 025

\bibitem[{Hooper \& Silk(2005)}]{Hooper2004}
Hooper, D., \& Silk, J. 2005, Phys. Rev., D71, 083503

\bibitem[{Hooper {et~al.}(2010)Hooper, Spolyar, Vallinotto, \&
  Gnedin}]{Hooper2010}
Hooper, D., Spolyar, D., Vallinotto, A., \& Gnedin, N.~Y. 2010, arXiv:1002.0005

\bibitem[{Hooper {et~al.}(2009{\natexlab{b}})Hooper, Stebbins, \&
  Zurek}]{Hooper2008}
Hooper, D., Stebbins, A., \& Zurek, K.~M. 2009{\natexlab{b}}, Phys. Rev., D79,
  103513

\bibitem[{Hooper {et~al.}(2004)Hooper, Taylor, \& Silk}]{Hooper2003}
Hooper, D., Taylor, J.~E., \& Silk, J. 2004, Phys. Rev., D69, 103509

\bibitem[{Hooper \& Zurek(2009)}]{Ref:annih9}
Hooper, D., \& Zurek, K.~M. 2009, Phys. Rev., D79, 103529

\bibitem[{Ibarra {et~al.}(2009)Ibarra, Ringwald, Tran, \& Weniger}]{decay6}
Ibarra, A., Ringwald, A., Tran, D., \& Weniger, C. 2009, JCAP, 0908, 017

\bibitem[{Ibarra \& Tran(2009)}]{Ibarra2008}
Ibarra, A., \& Tran, D. 2009, JCAP, 0902, 021

\bibitem[{Ibe {et~al.}(2009{\natexlab{a}})Ibe, Murayama, Shirai, \&
  Yanagida}]{decay7}
Ibe, M., Murayama, H., Shirai, S., \& Yanagida, T.~T. 2009{\natexlab{a}}, JHEP,
  11, 120

\bibitem[{Ibe {et~al.}(2009{\natexlab{b}})Ibe, Murayama, \& Yanagida}]{BWreso1}
Ibe, M., Murayama, H., \& Yanagida, T.~T. 2009{\natexlab{b}}, Phys. Rev., D79,
  095009

\bibitem[{Iglesias \& Rogers(1996)}]{OPAL}
Iglesias, C.~A., \& Rogers, F.~J. 1996, Astrophys. J., 464, 943

\bibitem[{Iocco(2008)}]{Iocco2008}
Iocco, F. 2008, Astrophys. J., 677, L1

\bibitem[{Iocco {et~al.}(2008)}]{DMfs3}
Iocco, F., {et~al.} 2008, Mon. Not. Roy. Astron. Soc., 390, 1655

\bibitem[{Ishiwata {et~al.}(2009)Ishiwata, Matsumoto, \& Moroi}]{decay4}
Ishiwata, K., Matsumoto, S., \& Moroi, T. 2009, Phys. Lett., B675, 446

\bibitem[{Kadota {et~al.}(2010)Kadota, Freese, \& Gondolo}]{Kadota2010}
Kadota, K., Freese, K., \& Gondolo, P. 2010, arXiv:1003.4442

\bibitem[{Kamionkowski {et~al.}(2010)Kamionkowski, Koushiappas, \&
  Kuhlen}]{Kamionkowski2010}
Kamionkowski, M., Koushiappas, S.~M., \& Kuhlen, M. 2010, Phys. Rev., D81,
  043532

\bibitem[{Kane {et~al.}(2002)Kane, Wang, \& Wells}]{Kane2001}
Kane, G.~L., Wang, L.-T., \& Wells, J.~D. 2002, Phys. Rev., D65, 057701

\bibitem[{{Kippenhahn} \& {Weigert}(1990)}]{kipp}
{Kippenhahn}, R., \& {Weigert}, A. 1990, {Stellar Structure and Evolution}
  (Springer-Verlag)

\bibitem[{Klypin {et~al.}(2010)Klypin, Trujillo-Gomez, \&
  Primack}]{2010arXiv1002.3660K}
Klypin, A., Trujillo-Gomez, S., \& Primack, J. 2010, arXiv:1002.3660

\bibitem[{Krauss {et~al.}(1985)Krauss, Freese, Press, \& Spergel}]{krauss}
Krauss, L.~M., Freese, K., Press, W., \& Spergel, D. 1985, Astrophys. J., 299,
  1001

\bibitem[{Lattanzi \& Silk(2009)}]{Som2}
Lattanzi, M., \& Silk, J.~I. 2009, Phys. Rev., D79, 083523

\bibitem[{{Lenzuni} {et~al.}(1991){Lenzuni}, {Chernoff}, \&
  {Salpeter}}]{lenzuni}
{Lenzuni}, P., {Chernoff}, D.~F., \& {Salpeter}, E.~E. 1991, \apjs, 76, 759

\bibitem[{Malyshev {et~al.}(2009)Malyshev, Cholis, \& Gelfand}]{Malyshev2009}
Malyshev, D., Cholis, I., \& Gelfand, J. 2009, Phys. Rev., D80, 063005

\bibitem[{March-Russell {et~al.}(2008)March-Russell, West, Cumberbatch, \&
  Hooper}]{Ref:MarchRusselletal08}
March-Russell, J., West, S.~M., Cumberbatch, D., \& Hooper, D. 2008, JHEP, 07,
  058

\bibitem[{{McKee} \& {Tan}(2008)}]{McKeeTan08}
{McKee}, C.~F., \& {Tan}, J.~C. 2008, \apj, 681, 771

\bibitem[{Meade {et~al.}(2010)Meade, Papucci, Strumia, \& Volansky}]{Meade2010}
Meade, P., Papucci, M., Strumia, A., \& Volansky, T. 2010, Nucl. Phys., B831,
  178

\bibitem[{Meade {et~al.}(2009)Meade, Papucci, \& Volansky}]{Meade2009}
Meade, P., Papucci, M., \& Volansky, T. 2009, JHEP, 12, 052

\bibitem[{Mertsch \& Sarkar(2009)}]{Mertsch2009}
Mertsch, P., \& Sarkar, S. 2009, Phys. Rev. Lett., 103, 081104

\bibitem[{Moskalenko \& Wai(2007)}]{moskalenko}
Moskalenko, I.~V., \& Wai, L.~L. 2007, Astrophys. J., 659, L29

\bibitem[{Navarro {et~al.}(1996)Navarro, Frenk, \& White}]{NFW}
Navarro, J.~F., Frenk, C.~S., \& White, S. D.~M. 1996, Astrophys. J., 462, 563

\bibitem[{Nelson \& Spitzer(2008)}]{B1}
Nelson, A.~E., \& Spitzer, C. 2008, arXiv:0810.5167

\bibitem[{Nomura \& Thaler(2009)}]{nomura}
Nomura, Y., \& Thaler, J. 2009, Phys. Rev., D79, 075008

\bibitem[{Omukai \& Nishi(1998)}]{OmuNis98}
Omukai, K., \& Nishi, R. 1998, Astrophys. J., 508, 141

\bibitem[{Pallis(2010)}]{Pallis2009}
Pallis, C. 2010, Nucl. Phys., B831, 217

\bibitem[{Profumo(2005)}]{Som1}
Profumo, S. 2005, Phys. Rev., D72, 103521

\bibitem[{Profumo(2008)}]{Profumo08}
---. 2008, arXiv:0812.4457

\bibitem[{{Ripamonti} \& {Abel}(2005)}]{RipAbe05}
{Ripamonti}, E., \& {Abel}, T. 2005, arXiv:astro-ph/0507130

\bibitem[{Ripamonti {et~al.}(2009)}]{DMfs4}
Ripamonti, E., {et~al.} 2009, PoS, IDM2008, 075

\bibitem[{Ripamonti {et~al.}(2010)}]{Ripamonti:2010ab}
---. 2010, arXiv:1003.0676

\bibitem[{{Ryden} \& {Gunn}(1987)}]{blumenthal3}
{Ryden}, B.~S., \& {Gunn}, J.~E. 1987, \apj, 318, 15

\bibitem[{{Salati} \& {Silk}(1989)}]{salatisi}
{Salati}, P., \& {Silk}, J. 1989, \apj, 338, 24

\bibitem[{{Scott} {et~al.}(2007){Scott}, {Edsj{\"o}}, \& {Fairbairn}}]{scott1}
{Scott}, P., {Edsj{\"o}}, J., \& {Fairbairn}, M. 2007, arXiv:0711.0991

\bibitem[{Scott {et~al.}(2008)Scott, Fairbairn, \& Edsjo}]{scott2}
Scott, P., Fairbairn, M., \& Edsjo, J. 2008, Mon. Not. Roy. Astron. Soc., 394,
  82

\bibitem[{Shaviv {et~al.}(2009)Shaviv, Nakar, \& Piran}]{Shaviv2009}
Shaviv, N.~J., Nakar, E., \& Piran, T. 2009, Phys. Rev. Lett., 103, 111302

\bibitem[{{Shirai} {et~al.}(2009){Shirai}, {Takahashi}, \& {Yanagida}}]{decay5}
{Shirai}, S., {Takahashi}, F., \& {Yanagida}, T.~T. 2009, Physics Letters B,
  675, 73

\bibitem[{Sivertsson \& Gondolo(2010)}]{Sivertsson2010}
Sivertsson, S., \& Gondolo, P. 2010, arXiv:1006.0025

\bibitem[{{Sommerfeld}(1931)}]{Ref:Somm31}
{Sommerfeld}, A. 1931, Annalen der Physik, 403, 257

\bibitem[{Spolyar {et~al.}(2009)Spolyar, Bodenheimer, Freese, \&
  Gondolo}]{DSnl}
Spolyar, D., Bodenheimer, P., Freese, K., \& Gondolo, P. 2009, Astrophys. J.,
  705, 1031

\bibitem[{Spolyar {et~al.}(2008)Spolyar, Freese, \& Gondolo}]{DS2}
Spolyar, D., Freese, K., \& Gondolo, P. 2008, Phys. Rev. Lett., 100, 051101

\bibitem[{Springel {et~al.}(2008)}]{2008MNRAS.391.1685S}
Springel, V., {et~al.} 2008, Mon. Not. Roy. Astron. Soc., 391, 1685

\bibitem[{Tan \& McKee(2004)}]{Tan2003}
Tan, J.~C., \& McKee, C.~F. 2004, Astrophys. J., 603, 383

\bibitem[{Taoso {et~al.}(2008)Taoso, Bertone, Meynet, \& Ekstrom}]{DMfs1}
Taoso, M., Bertone, G., Meynet, G., \& Ekstrom, S. 2008, Phys. Rev., D78,
  123510

\bibitem[{Tinker {et~al.}(2010)}]{Tinker2010}
Tinker, J.~L., {et~al.} 2010, arXiv:1001.3162

\bibitem[{Yin {et~al.}(2009)}]{B2}
Yin, P.-f., {et~al.} 2009, Phys. Rev., D79, 023512

\bibitem[{Yoon {et~al.}(2008)Yoon, Iocco, \& Akiyama}]{DMfs2}
Yoon, S.-C., Iocco, F., \& Akiyama, S. 2008, Astrophys. J., 688, L1

\bibitem[{Yoshida {et~al.}(2003)Yoshida, Abel, Hernquist, \&
  Sugiyama}]{Yoshidaetal03}
Yoshida, N., Abel, T., Hernquist, L., \& Sugiyama, N. 2003, Astrophys. J., 592,
  645

\bibitem[{Yoshida {et~al.}(2006)Yoshida, Omukai, Hernquist, \&
  Abel}]{Yoshidaetal06}
Yoshida, N., Omukai, K., Hernquist, L., \& Abel, T. 2006, Astrophys. J., 652, 6

\bibitem[{{Young}(1980)}]{young}
{Young}, P. 1980, \apj, 242, 1232

\bibitem[{Zackrisson {et~al.}(2010{\natexlab{a}})}]{Zackrisson2010}
Zackrisson, E., {et~al.} 2010{\natexlab{a}}, arXiv:1002.3368

\bibitem[{Zackrisson {et~al.}(2010{\natexlab{b}})}]{Zackrisson2010b}
---. 2010{\natexlab{b}}, arXiv:1006.0481

\bibitem[{Zhao {et~al.}(2009)Zhao, Jing, Mo, \& Boerner}]{Zhao2008}
Zhao, D.~H., Jing, Y.~P., Mo, H.~J., \& Boerner, G. 2009, Astrophys. J., 707,
  354

\bibitem[{Zhao {et~al.}(2003)Zhao, Jing, Mo, \& Borner}]{Zhao2003}
Zhao, D.-H., Jing, Y.~P., Mo, H.~J., \& Borner, G. 2003, Astrophys. J., 597, L9

\bibitem[{Zurek(2009)}]{Ref:annih6}
Zurek, K.~M. 2009, Phys. Rev., D79, 115002

\end{thebibliography}

\end{document}